\documentclass[11pt,preprint]{aastex6}  
\usepackage{hyperref}
\hypersetup{pdfauthor={Song et al.}, pdftitle={A2107}}
\usepackage{physics}
\usepackage{natbib}
\usepackage{subfigure}
\usepackage{url}
\usepackage{import}
\usepackage{multirow}
\usepackage{booktabs}

\newcommand{\fns}{\footnotesize}
\slugcomment{Last updated: \today}
\shorttitle{A2107}
\shortauthors{Song et al.}

\begin{document}

\title{A Redshift Survey of the Nearby Galaxy Cluster Abell 2107:
	Global Rotation of the Cluster and
	its Connection to Large-scale Structures in the Universe}

\author{Hyunmi Song\altaffilmark{1,2}, 
		Ho Seong Hwang\altaffilmark{3}, Changbom Park\altaffilmark{2}, 
		Rory Smith\altaffilmark{1},
		Maret Einasto\altaffilmark{4}}

\altaffiltext{1}{Korea Astronomy \& Space Science Institute, Daedeokdae-ro 776, Yuseong-gu, Daejeon 34055, Korea}
\altaffiltext{2}{School of Physics, Korea Institute for Advanced Study, Hoegiro 85, Dongdaemun-gu, Seoul 02455, Korea}
\altaffiltext{3}{Quantum Universe Center, Korea Institute for Advanced Study, Hoegiro 85, Dongdaemun-gu, Seoul 02455, Korea}
\altaffiltext{4}{Tartu Observatory, University of Tartu, Observatooriumi 1, 61602 T\~oravere, Estonia}

\begin{abstract}
We study the rotational motion of the galaxy cluster Abell 2107 at redshift $z=0.04$
	and its connection to nearby large-scale structures using a large amount of spectroscopic redshift data.
By combining 978 new redshifts from the MMT/Hectospec observations with data in the literature,
	we construct a large sample of 1968 galaxies with measured redshifts 
	at clustercentric radius $R<60^\prime$.
Our sample has high (80\%) and spatially uniform completeness 
	at $r$-band apparent magnitude $m_{r\textrm{,Petro,0}}<19.1$.
We first apply the caustic method to the sample and
	identify 285 member galaxies of Abell 2107 at $R<60^\prime$.
Then the rotation amplitude and the position angle of rotation axis are measured.
We find that the member galaxies show strong global rotation 
	at $R<20^\prime$ ($v_\textrm{rot}/\sigma_{v}\sim$0.6)
	with a significance of $>3.8\sigma$,
	which is confirmed by two independent methods.
The rotation becomes weaker in outer regions.
There are at least five filamentary structures 
	that are connected to the cluster
	and that consist of known galaxy groups.
These structures are smoothly connected to the cluster,
	which seem to be inducing the global rotation of the cluster
	through inflow of galaxies.
\end{abstract}

\keywords{cosmology: large-scale structure of universe 
	-- galaxies: clusters: individual (Abell 2107) -- galaxies: distances and redshifts -- galaxies: kinematics and dynamics 
	-- methods: observational
	-- surveys}

\section{INTRODUCTION}
Galaxy clusters are the largest gravitationally bound systems in the universe,
	which are often formed in places where large-scale filaments meet.
Matter accretes into a cluster through such filaments
	that are stretched to different directions.
As a result, there is usually no preferred direction in the motion of the matter in clusters,
	which means that, overall, cluster galaxies present random motions.
This is why clusters are generally considered as a pressure-supported system 
	rather than a rotation-supported system.
However, several observational studies found some clusters with non-negligible global rotation
	\citep{MH1983,OH1992,Tovmassian2002,Kalinkov_etal2005,HL2007,Hand_etal2012,Tovmassian2015,MP2017}.

The observed global rotation of clusters has been inferred from
	spatial segregation of velocity of cluster galaxies \citep{OH1992}, or
	velocity gradient with respect to either clustrocentric distance along an axis \citep{Kalinkov_etal2005}
	or position angle \citep{HL2007}.
However, some studies suggest that such observations could be explained by
	mechanisms other than cluster rotation.
One of them is the presence of substructures that overlap along the line-of-sight direction.
It is not easy to distinguish the two cases:
	a single rotating cluster vs. a system with two overlapping clusters.
Nevertheless, there are several clusters
	that show the spatially segregated velocity structure or 
	velocity gradient without strong evidence of subclustering,
	which is consistent with cluster rotation.
For example, Abell 2107 (A2107 hereafter), first claimed by \citet{OH1992} as a rotating cluster,
	shows a Gaussian velocity distribution, modest velocity dispersion, and a single number-density peak,
	which strongly suggest that A2107 is a single cluster in rotation.
\citet{HL2007} have identified six probable rotating clusters
	that show spatial segregation of galaxy velocities with a single number-density peak 
	from the sample of 899 clusters in the Sloan Digital Sky Survey (SDSS, \citealt{York_etal2000}). 
Although the number of rotating clusters is not so large,
	the global rotation can affect the cluster mass measurement
	that is usually based on the assumption of no rotation in clusters
	\citep[e.g.][]{LM1960,TW1986,Merritt1988,DG1997,Diaferio1999}
	and the measurement of the kinetic Sunyaev-Zel'dovich effect \citep[e.g.][]{CC2002,CM2002}.

There are several methods to measure the global rotation of clusters
	\citep{Kalinkov_etal2005,HL2007,MP2017}.
Each method has its own pros and cons,
	but all of them are limited by the projection effect;
	it is difficult to know the direction of the three-dimensional rotation axis.
To quantify well the global rotation of clusters under such limitation,
	it is important to use a dense sample of cluster galaxies \citep{Takizawa2000,MP2017}.
Moreover, a velocity field of a cluster is a combination of random motion
	and ordered rotational motion, and thus the signal of the rotational motion could be noisy.
This could be overcome through dense sampling as well.
We therefore observe the well-known relaxed, rotating cluster A2107
	to densely sample the galaxies around the cluster 
	at clustercentric radius $R<60^\prime$ (2.06$\,h^{-1}$Mpc).
The resulting sampling density of our survey by combining the data in the literature 
	is more than twice denser than the ones in previous studies.

Another important issue regarding the cluster rotation is to understand the origin of the rotation.
Theoretically, off-axis merging of two clusters \citep{Peebles1969,Ricker1998,Takizawa2000,RF2000}, 
	global rotation of the universe \citep{Li1998}, 
	and mass accretion from surrounding large-scale structures
	can be responsible for the acquired angular momentum of clusters.
The off-axis merging scenario could be examined by the study of merging history
	with X-ray data.
A2107 has been known to have a regular X-ray morphology 
	\citep[][see contours in Figure 1 of this study]{BT1996}
	without strong evidence of merging \citep{Fujita_etal2006,Lagana_etal2010}.
However, detailed analysis of the X-ray data (e.g. temperature map)
	is limited only to the central region ($R\lesssim10^\prime$).
This makes it difficult for us to fully understand the kinematics 
	of the intracluster medium (ICM) of A2107,
	which is crucial for predicting the merging history.
On the other hand, if the universe is globally rotating,
	it is expected that celestial objects obtain angular momentum upon their formation
	\citep{Li1998,Godlowski_etal2003}.
This theory predicts a correlation between the angular momentum and the mass of the system.
To test the prediction from this theory about cluster rotation 
	(i.e. rotation induced by the global rotation of the universe),
	we need data for many clusters over a wide range of masses; 
	this requires a systematic survey of clusters.
The third scenario for the origin of the cluster rotation 
	(i.e. mass accretion from surrounding large-scale structures)
	could be tested with wide-field spectroscopic survey data.
Thanks to large surveys 
	including SDSS and 2dF Galaxy Redshift Survey (2dFGRS, \citealt{Colless_etal2001}), 
	we are in a good position to investigate this possibility.

In this study, we aim to better measure the global rotation of A2107
	using the dense sample of galaxies from our redshift survey.
With a focus on the mass accretion from large-scale structures
	as a possible origin of the cluster rotation,
	we will examine velocity fields of the large-scale structures around A2107
	and their connections to the cluster using the combined data of the SDSS and our MMT/Hectospec observations.
We also search for counterparts of the structures among known galaxy groups and clusters in SDSS.

This paper is part of the KIAS redshift survey of nearby galaxy groups and clusters,
	which aims to construct dense samples of group/cluster galaxies
	for the studies of kinematics and stellar populations
	\citep{Song_etal2017,Deshev_etal2017,Park_etal2018}.
This paper is constructed as follow. 
In Section \ref{sec-data}, we describe the observational data
	obtained from the MMT/Hectospec redshift survey.
We determine the cluster membership in Section \ref{sec-member},
	and present the results from the analysis of global rotation of A2107 in Section \ref{sec-rotation}.
We discuss the origin of angular momentum in the cluster in Section \ref{sec-discuss},
	and summarize the results in Section \ref{sec-summary}.
Throughout, we adopt flat $\Lambda$CDM cosmological parameters:
	$H_0 = 100$ $h$ km s$^{-1}$ Mpc$^{-1}$, 
	$\Omega_{\Lambda}=0.7$, and $\Omega_{m}=0.3$.
All quoted errors in measured quantities are 1$\sigma$.

\section{DATA}\label{sec-data}
\subsection{Photometric Data}\label{sec-data-phot}
We select targets for spectroscopic observations 
	from the photometric data of the SDSS data release 12 \citep[DR12;][]{SDSSDR12}.
The goal of the spectroscopic survey in this study is to cover the large area around A2107
	with high completeness and uniformity.
To achieve this goal, we select galaxies without measured redshifts in the literature 
	from the photometric data
	(i.e. extended sources based on the probPSF parameter in the SDSS database
	; see Section 4.2 of \citealt{Strauss_etal2002} for the star-galaxy separation in detail),
	and impose no selection criteria other than magnitude;
	we assign higher priority to brighter galaxies 
	in $r$-band apparent magnitude ($m_{r\textrm{,Petro,0}}$).\footnote{
		The first and second subscripts denote a photometric band 
		and a model used to measure magnitudes, respectively.
		The subscript 0 of magnitudes represents magnitudes after the Galactic extinction correction.}

\subsection{Spectroscopic Observations and Data Reduction}\label{sec-data-spec}
\begin{figure*}
\center
\includegraphics[width=0.9\textwidth]{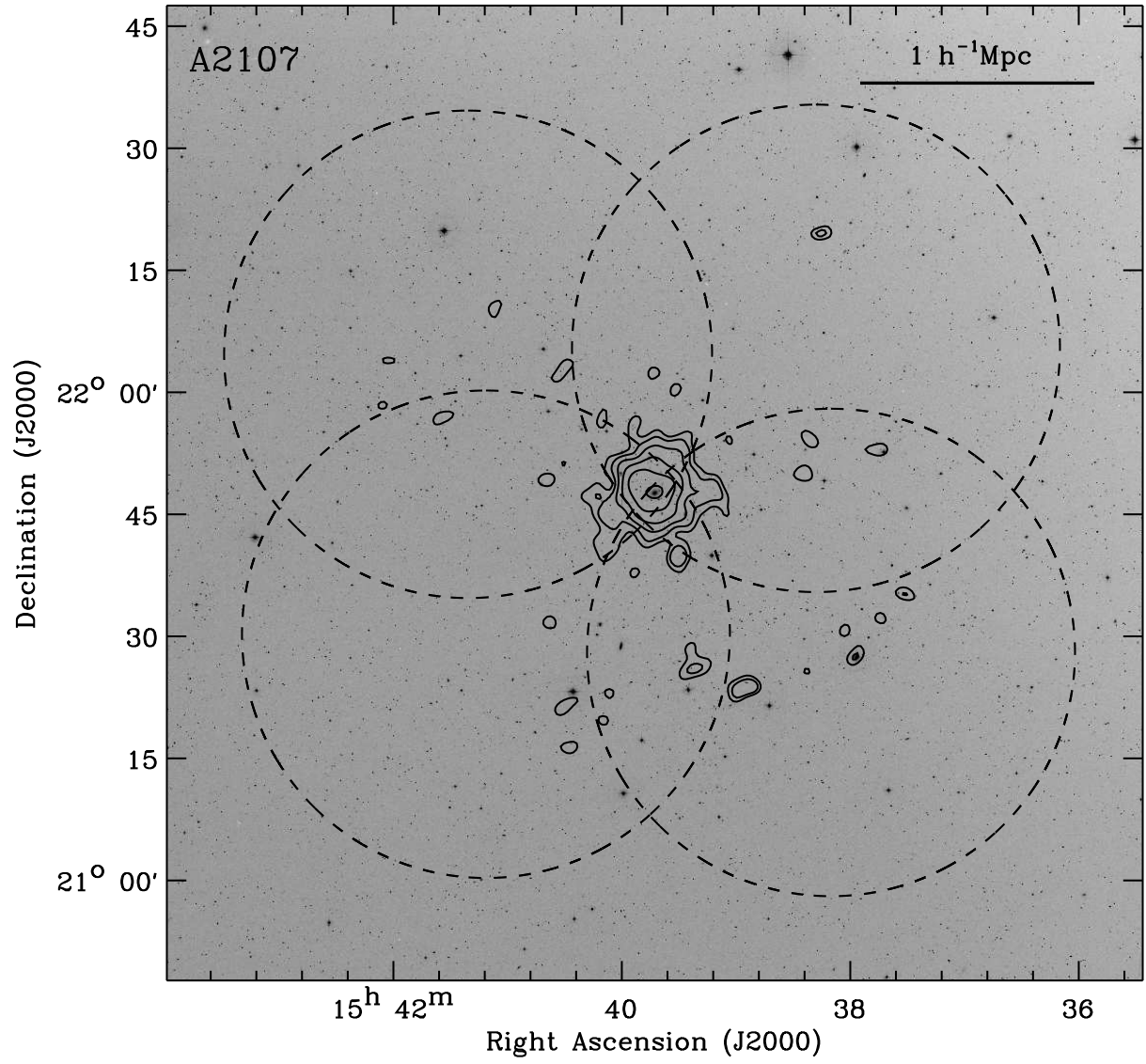}
\caption{MMT/Hectospec target fields (dashed circles)
	and X-ray contours (solid lines) overlaid on the
	optical Digitized Sky Survey.
	The X-ray image is taken from Einstein/IPC.
}\label{fig-xray}
\end{figure*}

We use the Hectospec installed on the MMT 6.5m telescope for spectroscopic observations 
	\citep{Fabricant_etal1998,Fabricant_etal2005}.
The Hectospec is a 300 fiber multi-object spectrograph with a circular field of view (FOV)
	of $1^\circ$ diameter.
We use the 270 line mm$^{-1}$ grating of Hectospec
    that provides a dispersion of 1.2$\AA$ pixel$^{-1}$ and a resolution of $\sim$6$\AA$.
We observe four fields with 3$\times$15 (or 20) minute exposure each,
    and obtained spectra covering the wavelength range 3500--9150$\AA$.
To cover the large area around A2107,
	we positioned the four fields around the cluster as shown in Figure \ref{fig-xray} (dashed circles).
The positions of the four MMT/Hectospec fields and the numbers of targets are summarized in Table \ref{tab-obs}.

We reduce the Hectospec spectra with HSRED v2.0,
    which is an updated reduction pipeline originally developed by Richard Cool.
We then use RVSAO \citep{Kurtz_Mink1998} to determine the redshifts
    by cross-correlating the spectra with templates.
RVSAO gives the \citet{Tonry_Davis1979}'s $r$-value for each spectrum,
    which is an indicator of cross-correlation reliability;
    we select only those galaxies with $r>4$,
    consistent with the limit confirmed by visual inspection
    \citep{SHELS2014,SHELS2016}.
In the end, we obtain 978 reliable redshifts from the observations of 1023. 
The main reason why we could not determine the reliable redshifts for the 45 objects
	is because of their spectra with low signal-to-noise ratios.
We combine these data with those from the SDSS DR12 \citep{SDSSDR12} and from the literature.
In total, we have 1968 redshifts at $R<60^\prime$:
    978 from our MMT/Hectospec observations, 
	977 from the SDSS DR12, 
	five from \citet{Cava_etal2009},
	three from \citet{OH1992},
	three from \citet{Smith_etal2004},
	one from \citet{White_etal2000}, and
	one from the FAST Spectrograph Archive.\footnote{
	\href{http://tdc-www.harvard.edu/cgi-bin/arc/fsearch}{http://tdc-www.harvard.edu/cgi-bin/arc/fsearch}}

\begin{figure*}
\center
\includegraphics[width=0.9\textwidth]{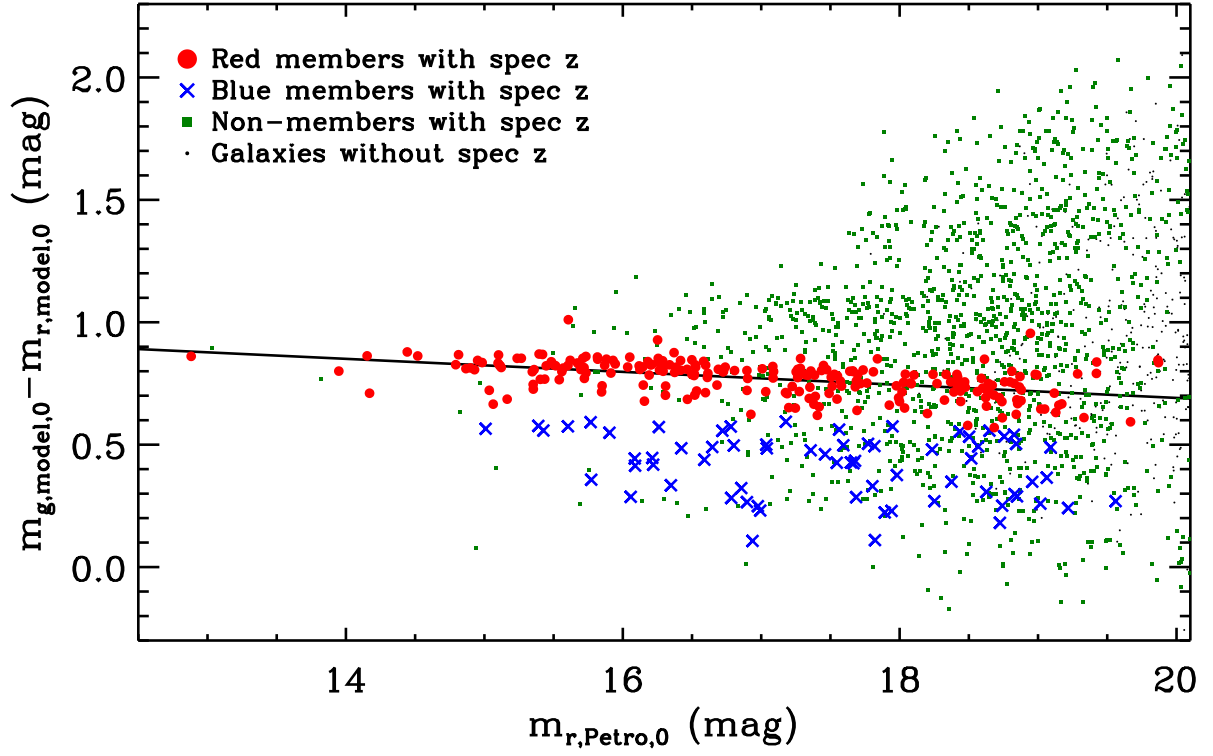}
\caption{$(g-r){-}r$
	color-magnitude diagram of A2107 galaxies at $R<60^\prime$.
Colored symbols are for galaxies with spectra.
Black dots represent galaxies without spectra,
	and we display only 10\% of the data for clarity.
Red dots and blue crosses are red and blue cluster member galaxies, respectively.
The red sequence (Equation (\ref{eqn_redsq})) is denoted by a black solid line.
}\label{fig-cmr}
\end{figure*}

Figure \ref{fig-cmr} shows a $(g-r){-}r$ color-magnitude distribution of the objects at $R<60^\prime$.
Galaxies with measured redshifts are denoted by symbols in color,
	while those without spectra are denoted by black dots.
Because we have not imposed any selection criteria
	for the targets of spectroscopic observations (See Section \ref{sec-data-phot}),
	colored symbols (red filled circles, blue crosses, and green squares)
	cover the whole color range without having any exclusion region.
Their distribution clearly shows the red sequence of A2107.
We use the cluster member galaxies (see Section \ref{sec-member} for the member selection)
	to determine the best-fit relation of the red sequence, which is
\begin{equation}
	m_{g\textrm{,model,0}} - m_{r\textrm{,model,0}} 
	= 1.223 - 0.027\,m_{r\textrm{,Petro,0}}
\label{eqn_redsq}
\end{equation}
	(solid black line in Figure \ref{fig-cmr}).
The rms scatter around this relation is 0.056 mag.
We divide cluster member galaxies into red (red dots) and blue (blue crosses) populations
	using the line $3\sigma$ blueward of the best-fit red-sequence relation
	\citep{SanchezBlazquez_etal2009,Rines_etal2013,A383,Hwang_etal2014}.
Green squares are foreground or background galaxies.

\begin{figure*}
\center
\includegraphics[width=0.9\textwidth]{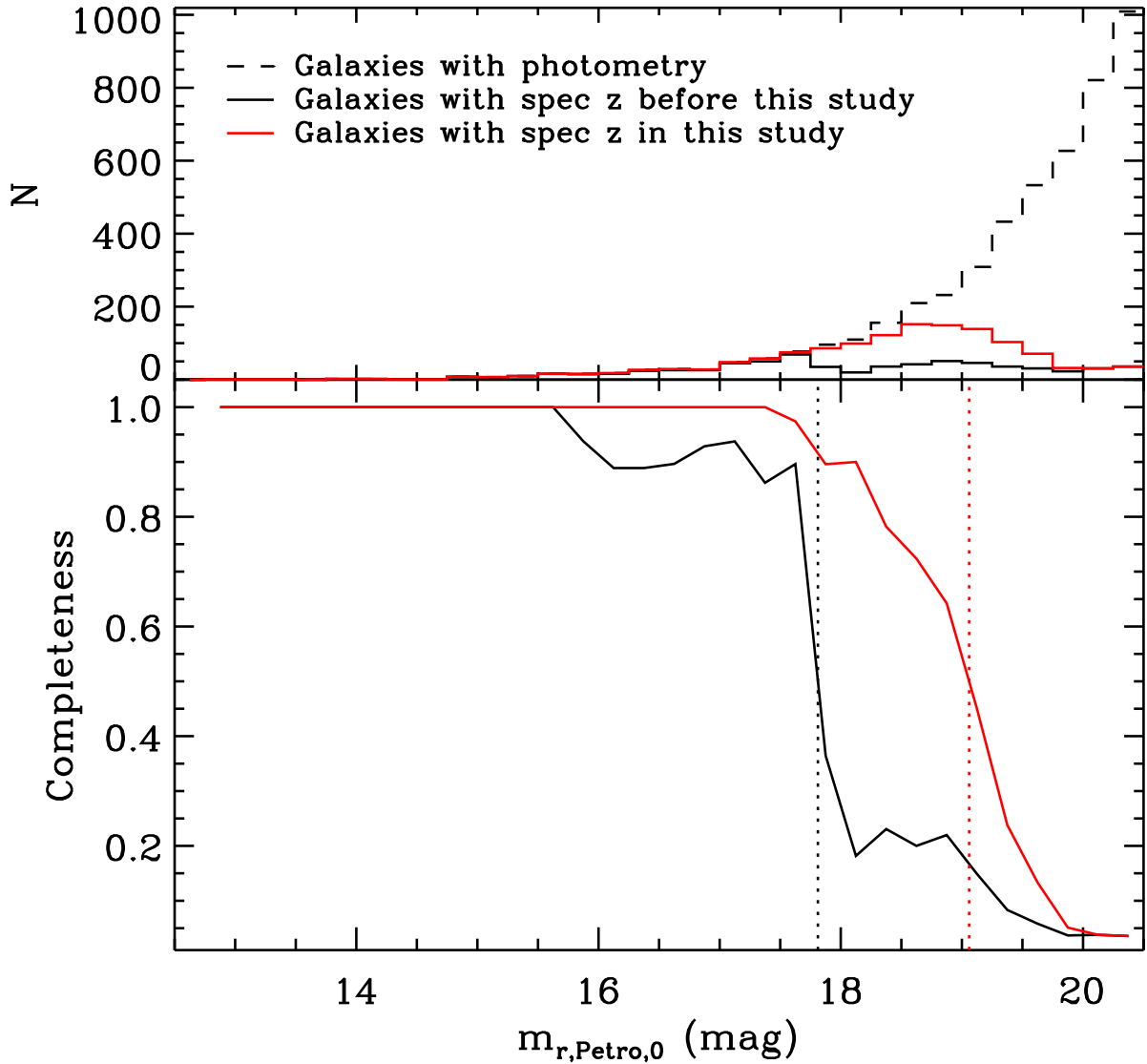}
\caption{(Top) Number of galaxies in the field of A2107 ($45^\prime\times45^\prime$) 
	as a function of $r$-band apparent magnitude:
	dashed line for all galaxies regardless of redshift measurement,
	black and red solid lines for galaxies with spectra 
	before and after our MMT/Hectospec observations, respectively.
(Bottom) Differential spectroscopic completeness.
Black and red solid lines are the completeness 
	before and after our MMT/Hectospec observations, respectively.
Vertical dotted lines at $m_{r\textrm{,Petro,0}}=17.8$ and 19.1
	denote the magnitude where the completenesses drops below 50\%. 
}\label{fig-comp-mag}
\end{figure*}

\begin{figure*}
\center
\includegraphics[width=0.9\textwidth]{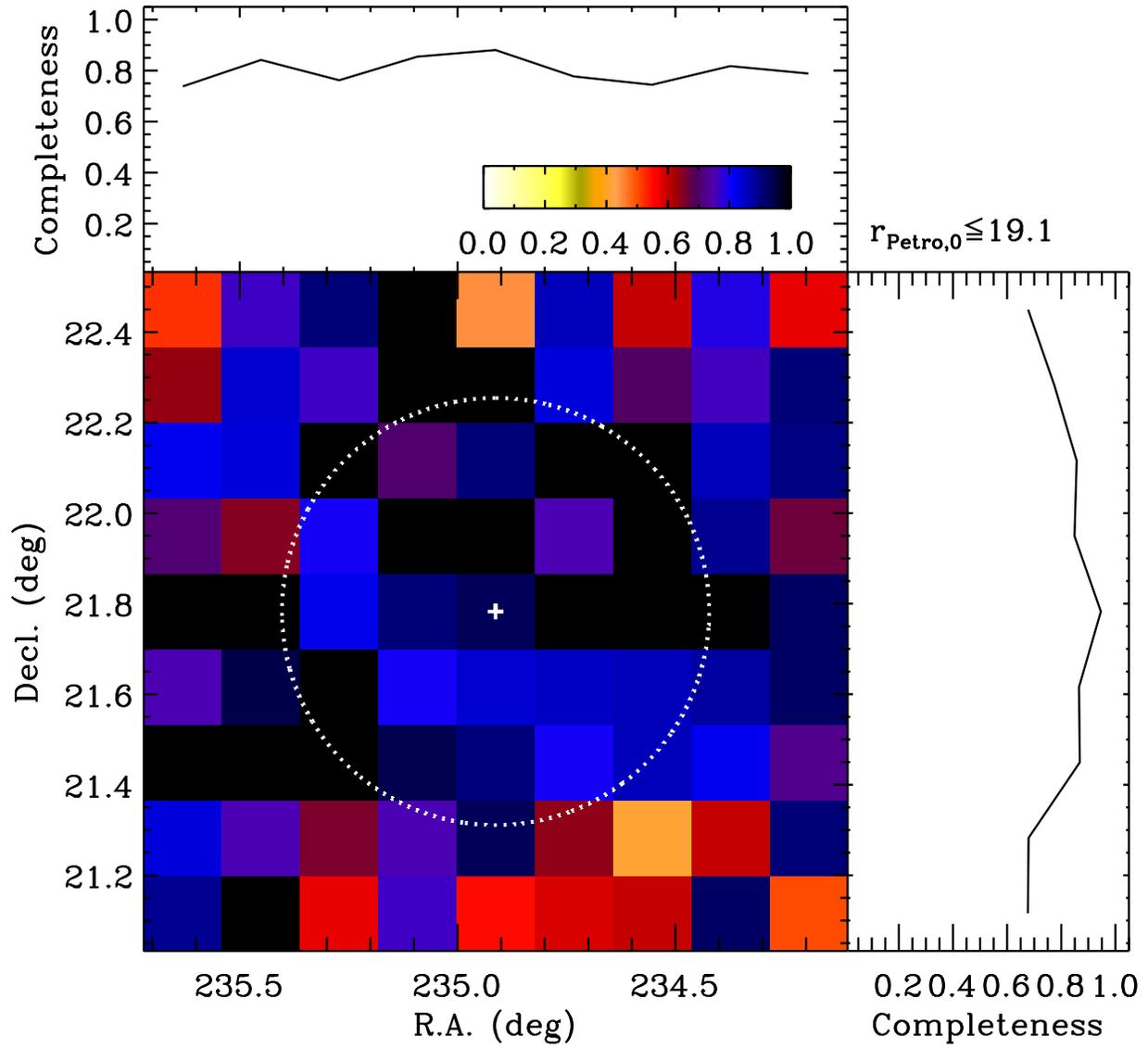}
\caption{Cumulative spectroscopic completeness 
	at $m_{r\textrm{,Petro,0}}\le19.1$
	in the field of A2107 ($45^\prime\times45^\prime$) 
	as a function of R.A. and decl. (bottom left).
The cumulative completenesses integrated over the decl. (top) 
	and R.A. (bottom right) ranges, respectively.
The white dotted circle denotes the virial radius of the cluster.
}\label{fig-comp-radec}
\end{figure*}

Figure \ref{fig-comp-mag} shows how complete the redshift data in the field of A2107 ($45^\prime\times45^\prime$) 
	are before and after our MMT/Hectospec survey as a function of $r$-band apparent magnitude.
In the top panel, the black dashed line denotes the number of galaxies regardless of redshift measurements,
	and the black and red solid lines denote the numbers of galaxies with measured redshifts 
	before and after this survey, respectively.
The bottom panel shows the spectroscopic completenesses before (black) and after (red) the survey, 
	which are calculated by the numbers of measured redshifts (i.e. solid lines in the top panel) 
	divided by the number of extended objects (i.e. dashed line in the top panel).
The vertical lines indicate the magnitude where the completeness starts to drop below 50\%.
That magnitude limit is increased more than one magnitude by including our new redshift data
	(from $m_{r\textrm{,Petro,0}}<17.8$ to $m_{r\textrm{,Petro,0}}<19.1$).
The cumulative completeness at $m_{r\textrm{,Petro,0}}<19.1$ is also increased 
	significantly from 60\% to 80\%.
The bottom-left panel of Figure \ref{fig-comp-radec} shows 
	the cumulative completeness at $m_{r\textrm{,Petro,0}}<19.1$ as functions of R.A. and decl.
The top and right panels display the integrated completenesses over the decl. and R.A. ranges, respectively.
All the three panels show that our redshift survey uniformly covers the cluster as planned.

\begin{figure*}
\center
\includegraphics[width=0.9\textwidth]{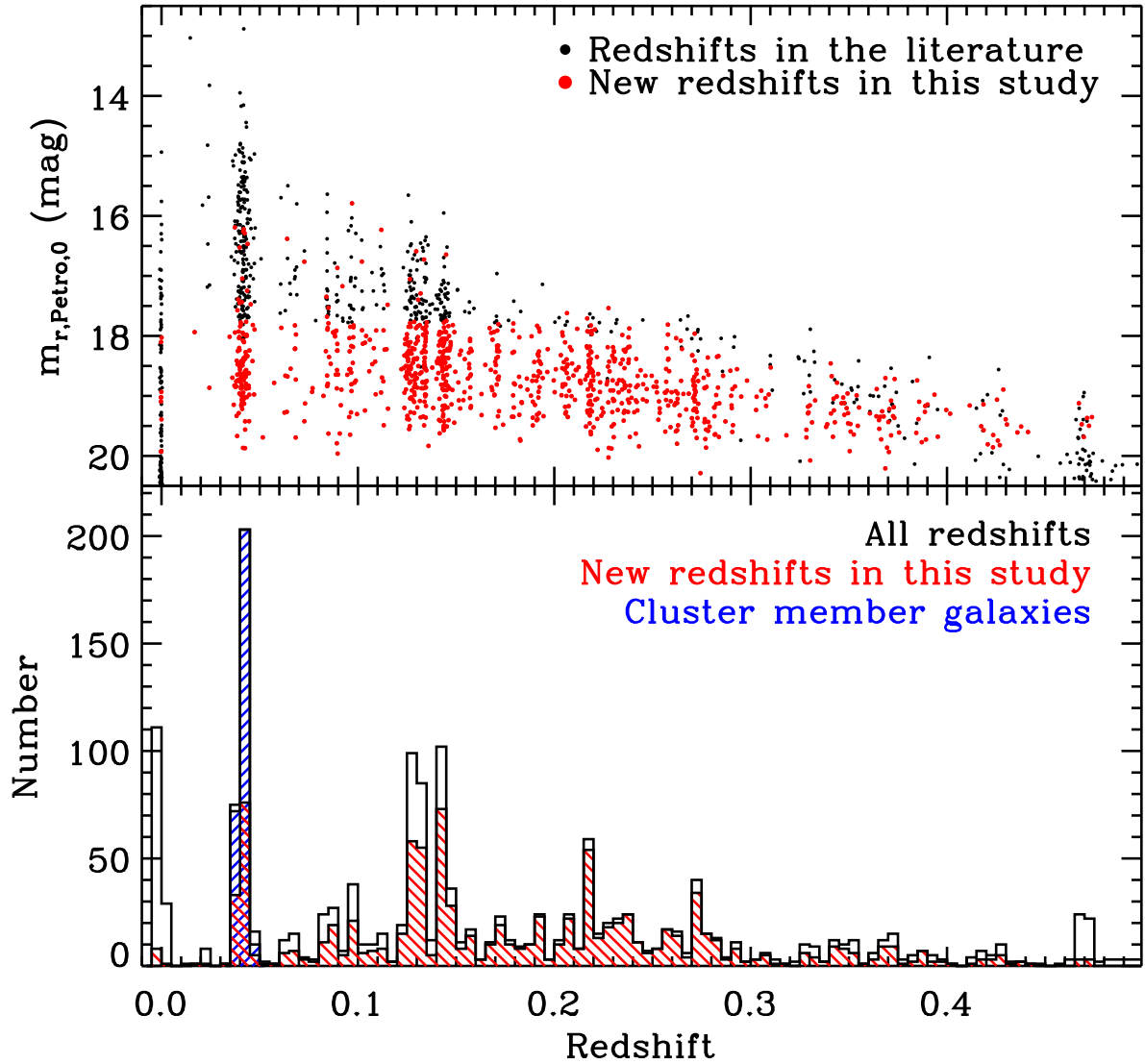}
\caption{(Top) Distribution of $r$-band apparent magnitude 
	of galaxies at $R<60^\prime$ as a function of redshift.
Black dots represent galaxies with spectra in the literature.
Red dots denote galaxies with new redshifts in this study.
(Bottom) Redshift histograms of galaxies at $R<60^\prime$.
Black histogram is for all galaxies with spectra
	and red one is for galaxies with new redshifts in this study.
Blue one shows the redshift distribution of member galaxies of A2107.
}\label{fig-zhist}
\end{figure*}

Figure \ref{fig-zhist} displays the distributions of galaxies as a function of redshift.
The top panel shows $r$-band apparent magnitudes of galaxies.
Black and red dots represent galaxies with measured redshifts 
	from the literature and our MMT/Hectospec survey, respectively.
Most targets in our survey are in the magnitude range of $m_{r\textrm{,Petro,0}}=18.0$--$19.5$,
	and many of them are background objects behind the cluster 
	(i.e. red histogram in the bottom panel).
The black one shows the total redshift distribution,
	while the blue one shows the distribution for the cluster member galaxies around $z=0.04$.
Interestingly, the histograms shows several peaks other than A2107 
	corresponding to other galaxy groups, but we focus only on A2107 in this study.

Table \ref{tab-cat} 
	lists 1968 redshifts in the central region of A2107 at $R<60^\prime$.
The table includes SDSS DR12 PhotObjID, R.A., decl., $r$-band apparent magnitude,
	the SDSS probPSF parameter, redshift, and its error, redshift source, 
	and cluster membership (see Section \ref{sec-member}).

\section{CLUSTER MEMBERSHIP WITH THE CAUSTIC TECHNIQUE}\label{sec-member}
Motions of galaxies bound in cluster gravitational potential well
	shape their distribution in a phase space of the radial velocity and projected clustercentric radius,
	resulting in a characteristic trumpet-shaped pattern in relaxed clusters.
The edges of this distribution, so-called caustics, correspond to the escape velocity at each radius
	\citep{Kaiser1987,RG1989,Diaferio_Geller1997}.
\citet{Diaferio_Geller1997} and \citet{Diaferio1999} developed the caustic technique, 
	which defines the trumpet-shaped pattern in the phase space
	and separates cluster galaxies from foreground and background galaxies.
Briefly, the technique smooths the density distribution in the phase space with an adaptive kernel,
	and finds the location of the caustics where the density reaches a certain threshold
	\citep[see][for more information]{Diaferio1999,Serra_etal2011}.
\citet{SD2013} demonstrated that the caustic technique works well in identifying cluster member galaxies
	within $3R_{200}$. 
They used 100 mock clusters from a cosmological $N$-body simulation, 
	showing that the technique recovers 95$\pm$3\% of true members.
We use The Caustic App v1.2, 
	an open source application developed by Anna Laura Serra and Antonaldo Diaferio,
	to apply this technique to the sample of galaxies with measured redshifts at $R<70^\prime$
	for the identification of the member galaxies in A2107.

\begin{figure*}
\center
\includegraphics[width=0.9\textwidth]{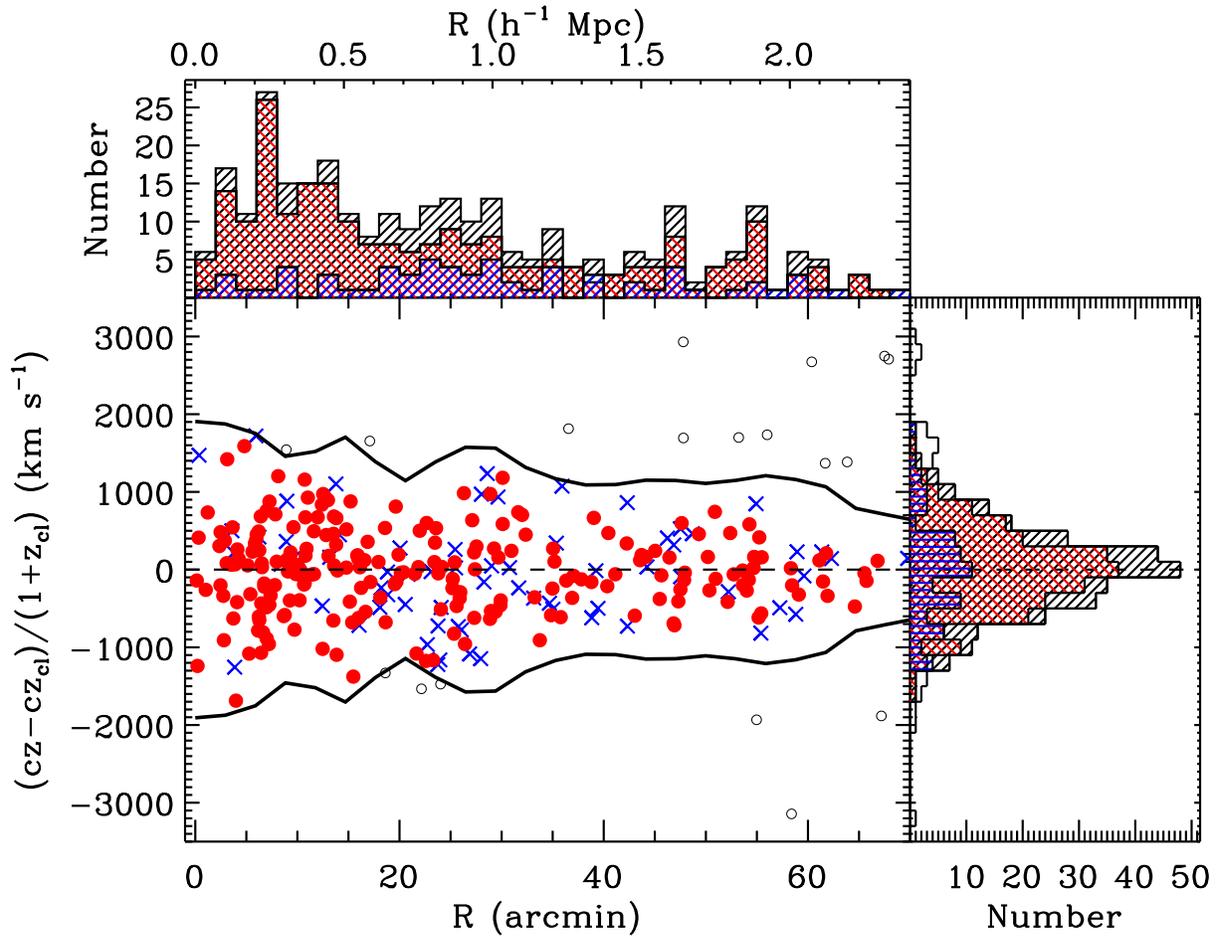}
\caption{(Bottom left) Clustercentric velocities of galaxies
	as a function of projected clustercentric radius.
Black thick lines represent the caustics.
Member galaxies are identified as those within the caustics.
Red dots and blue crosses are red and blue members, respectively.
Non-member galaxies outside the caustics are denoted by black open circles.
Histograms of clustercentric radii (top) and of clustercentric velocities (middle right)
	are also presented.
Black, red, and blue hatched histograms are for all, red, and blue members, respectively.
Black plane histogram is for galaxies including both members and non-members.
}\label{fig-caustic}
\end{figure*}

Figure \ref{fig-caustic} shows the distribution of galaxies in the phase space
	of the clustercentric velocity along the line of sight and clustercentric radius
	with the determined caustics (solid black lines) in the bottom left panel.
The caustics appears as the well-known trumpet shape and agrees well with the boundaries based on a visual impression.
This results in 285 member galaxies at $R<60^\prime$ ($\sim2.06\,h^{-1}$Mpc).
We further divide the member galaxies into two groups (red and blue)
	based on the positions in the color-magnitude diagram 
	(see Figure \ref{fig-cmr} and Section \ref{sec-data-spec}).
The top panel shows histograms of red members (red), blue members (blue), 
	and all the galaxies (black)
	as a function of clustercentric radius.
Similarly, the right panel shows those as a function of the clustercentric velocity.
As expected, the red members dominate the cluster population 
	and are mostly found in the inner region of the cluster.

The caustic technique also provides a position of the cluster center 
	(i.e. hierarchical center; see Appendix A of \citealt{Diaferio1999} for more details) 
	and a mass profile.
The hierarchical center of A2107 determined from this technique is
	R.A.$=234^\circ.986047$, decl.$=21^\circ.763402$, and $z=0.041410$,
	which is roughly consistent with the position of cD galaxy
	(UGC 09958, R.A.$=234^\circ.91269$, decl.$=21^\circ.78272$, and $z=0.0418939$).
We use the hierarchical center as a cluster center
	for the analysis of cluster rotation from now on.
We adopt $r_{200}\sim1.01\,h^{-1}$Mpc, 
	which is obtained from the fit of the caustic mass profile
	with the Navarro-Frenk-White profile \citep{NFW1996}
	along with $M_{200}\sim2.89\times10^{14}h^{-1}M_{\odot}$.

\section{MEASUREMENTS OF GLOBAL ROTATION OF A2107}\label{sec-rotation}
\subsection{The Dressler-Shectman Test and Spatial Segregation of Velocity Distribution}\label{sec-OH}
There are several studies that suggest the global rotation of A2107.
\citet{OH1992} showed that the radial velocities 
	of 68 cluster member galaxies are spatially correlated (see their Figures 5 and 6), 
	while there is no strong evidence for two distinct clusters aligned with the line of sight
	(single-peaked galaxy distribution, regular X-ray morphology, and Gaussian velocity distribution).
\citet{Kalinkov_etal2005}, using the same data, quantified the spatial correlation of velocities of cluster members
	by measuring the correlations between position angles and velocities in several radial ranges.
They found that the rotational effect is strong at clustercentric radius of 0.26--0.54 Mpc,
	and derived a rotation-corrected cluster mass.
Here, we first follow the analyses of \citet{OH1992} with newly added cluster members,
	and then quantify the rotation using two methods in the following section.
The improvements in this study are the use of members robustly determined from the caustic technique
	and the number of members increased by more than a factor of two (in the same R.A. and decl. ranges) 
	thanks to our MMT/Hectospec survey and the SDSS data.

\begin{figure*}
\center
\includegraphics[width=0.8\textwidth]{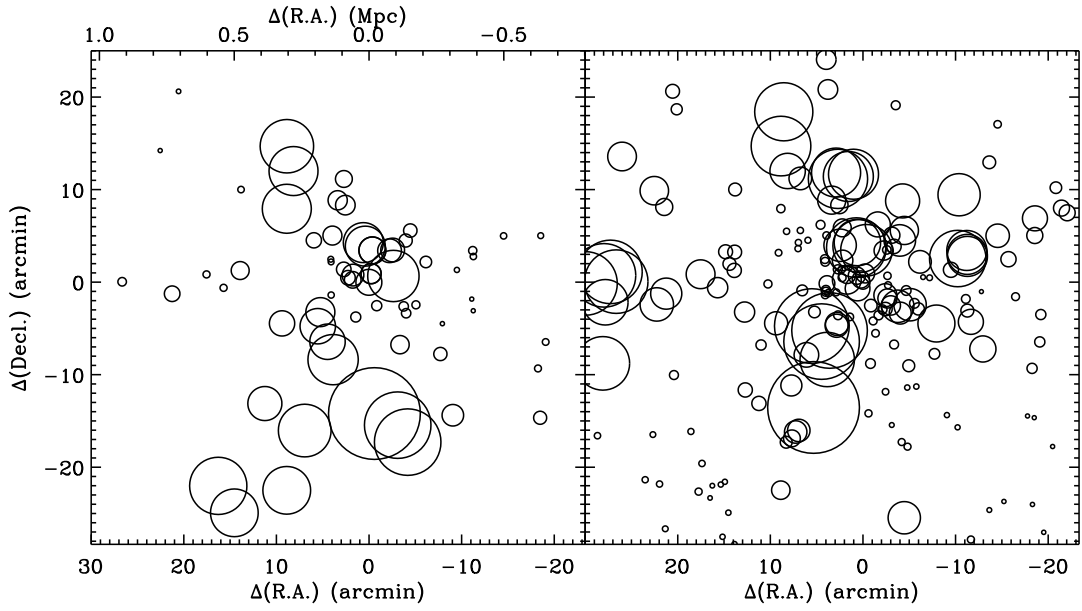}
\caption{Spatial distributions of member galaxies.
Each galaxy is presented by an open circle of which diameter is proportional to $e^\delta$.
$\delta$ denotes the difference between local (defined by 11 nearest neighbors) 
	and global (cluster) kinematics (Equation (\ref{eqn-ds})).
The left panel is for member galaxies in \citet[][see also their Figure 5]{OH1992},
	and the right panel is for member galaxies in this study.
Note that R.A. and decl. ranges are chosen to be identical to those in \citet{OH1992}
	for easy comparison.
}\label{fig-ds}
\end{figure*}

Figure \ref{fig-ds} shows the Dressler-Shectman (DS) diagrams \citep{DS1988}
	for the old data used in \citet[][left panel]{OH1992}
	and the new data in this study (right panel).
Each galaxy is represented by an open circle of which size is proportional to $e^\delta$.
The $\delta$ represents the difference between local kinematics and the global one,
	which is expressed as
\begin{equation}
	\delta^2 = (N_\textrm{nn}/\sigma_\textrm{cl}^2) 
			\left[ (\bar{v}_\textrm{local}- \bar{v}_\textrm{cl})^2 
				+ (\sigma_\textrm{local}-\sigma_\textrm{cl})^2 \right]
	\label{eqn-ds}
\end{equation}
	where 
	$\bar{v}_\textrm{local}$ and $\sigma_\textrm{local}$ are the local velocity mean and dispersion
	of $N_\textrm{nn}$ neighboring galaxies,
	and $\bar{v}_\textrm{cl}$ and $\sigma_\textrm{cl}$ are the global values of the cluster members 
	within $r_{200}$.
Following \citet{DS1988} we use $N_\textrm{nn}=11$ that includes a target galaxy and its 10 nearest neighbors.
The R.A. and decl. ranges of the figure are chosen to be the same as Figure 5 of \citet{OH1992}, 
	which roughly corresponds to the range of $R<r_{200}$.
Groups of large circles in the DS diagram indicate candidates of substructures 
	of which kinematics is not assimilated into the cluster yet.
While there are groups of large circles in common between the two panels,
	the right panel shows more subclustering features in general;
	there are substructures in the east and west not seen in the left panel,
	and the substructures in the northwest and southeast near the cluster center
	appear more prominently.
The subclustering features in the south significantly differ between the two panels.
The comparison of the DS diagrams between different galaxy samples
	suggests that detailed features of the DS diagram can change significantly
	depending on sampling density.
To quantify the significance of substructures,
	we run 1000 Monte Carlo simulations by randomly shuffling the measured redshifts among member galaxies,
	and count the number of simulations that have the cumulative deviation $\Delta=\Sigma_i \delta_i$
	larger than the one from the real data set.
We find that 1.8\% of the simulated clusters has such $\Delta$,
	indicating the significance of substructures of A2107 is high;
	\citet{OH1992} found 0\%, consistent with this study.
The DS tests in both \citet{OH1992} and this study suggest that
	A2107 contains significant velocity substructures.

\begin{figure*}
\center
\includegraphics[width=\textwidth]{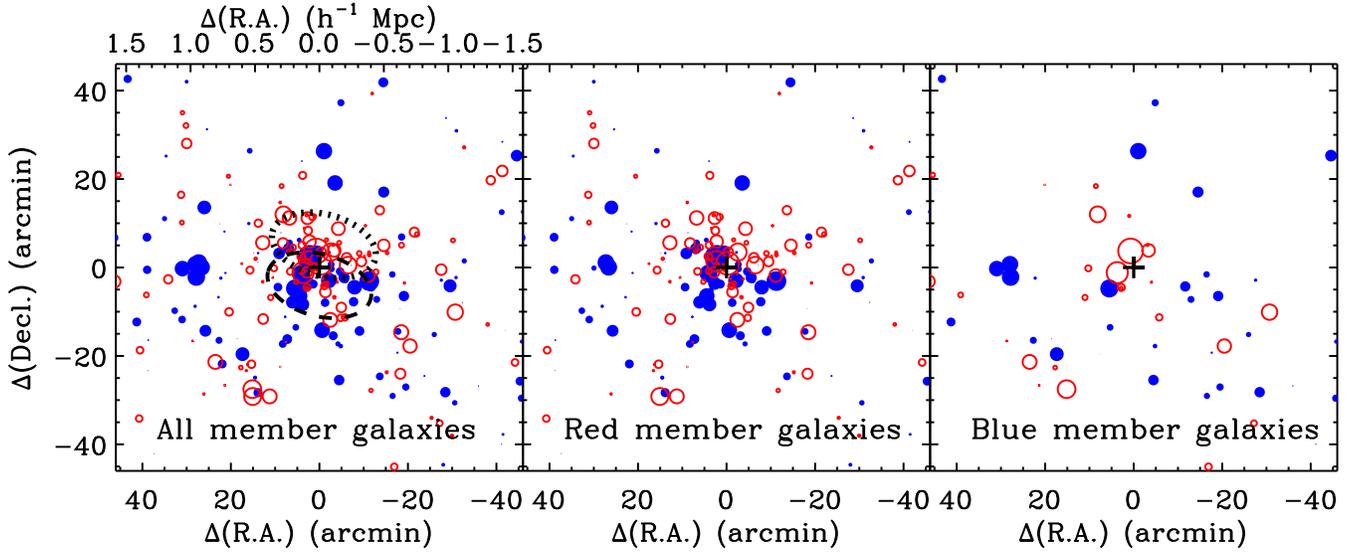}
\caption{Spatial distributions of member galaxies
	with their clustercentric velocities.
Galaxies are presented with red open circles (blue filled circles)
	when they have positive (negative) velocities.
Circle size is proportional to the magnitude of the velocity.
Plots are made separately for all members (left), red members (middle), and blue members (right).
The black dotted and dashed ovals in the left panel
	denote the receding group in the northwest
	and the preceding group in the southeast, respectively, at $R<20^\prime$.
}\label{fig-vspat}
\end{figure*}

To directly examine whether the global rotation of A2107 exists or not,
	we plot the spatial distribution of member galaxies in Figure \ref{fig-vspat}.
The galaxies with positive and negative clustercentric radial velocities
	are indicated by open red and filled blue circles, respectively.
The size of the circle is proportional to the clustercentric radial velocity.
The plot clearly shows that the galaxies with positive and negative velocities
	are spatially segregated at $R<20^\prime$ in the northwest and southeast
	(denoted by black dotted and dashed ovals), respectively.
This is consistent with the subclusterings seen in the DS diagram (right panel of Figure \ref{fig-ds}).
This segregation also appears when we plot red (middle panel) 
	and blue (right panel) members separately.
This spatial segregation of the galaxies with positive and negative velocities 
	was also seen in Figure 6 of \citet{OH1992}, consistent with our result.

\subsection{Measurements on the Rotation Axis and Amplitude of Cluster Galaxies}\label{sec-rotmeasure}

We now quantify the rotation of A2107 using two different methods suggested by 
	\citet[][hereafter HL07]{HL2007} and \citet[][hereafter MP17]{MP2017}.
The HL07 method is to fit the line-of-sight velocities of galaxies ($v_p$)
	with a sine function of position angle ($\theta$).
We rewrite their Equation (1) in the cluster rest-frame
	and add a minus sign in the right hand side
	to have a velocity component toward us in the east of the rotation axis 
	when the rotation axis points to the north:\footnote{
		It should be noted that position angle increases counterclockwise
		and the clustercentric velocity increases backward of the cluster.} 
\begin{equation}
	(v_\textrm{p}-v_\textrm{cl})/(1+z_\textrm{cl})
	= (cz-cz_\textrm{cl})/(1+z_\textrm{cl}) 
	= -v_\textrm{rot}\,\textrm{sin}(\theta-\theta_0)
	\label{eqn-vfit}
\end{equation}
	where $v_\textrm{cl}$ and $z_\textrm{cl}$ are the line-of-sight velocity
	and redshift of the cluster.
There are two fitting parameters, $v_\textrm{rot}$ and $\theta_0$,
	representing the rotation velocity 
	and the position angle of the rotation axis, respectively.
We use the notation in which $v_\textrm{rot}$ has always a positive value and
	$\theta_0$ is in the range $-180^\circ<\theta_0<180^\circ$.

Another method to determine the rotation axis and amplitude is suggested by MP17.
They divide the cluster galaxies into two samples (namely 1 and 2) 
	by the line going through the cluster center.
They then measure the difference of the mean galaxy velocities of the two samples,
	$\langle v_1\rangle-\langle v_2\rangle$.
They repeat this measurement by changing the position angles of galaxies by $\theta$.
Equivalently, one can rotate the division line of the two galaxy samples 
	by angle $\theta$ instead of galaxies.
We take this approach with $\theta$ to be the position angle of the division line.
In the result, they can obtain the mean velocity difference as a function of the angle $\theta$; 
	the velocity difference reaches its maximum (or minimum)\footnote{
		Being maximum or minimum depends on how to number the subsamples;
		we call the east side of the division line sample 1,
		and in that way the velocity difference reaches its minimum
		when the division line coincides with the rotation axis.}
	when the angle coincides with the position angle of the rotation axis.
They demonstrated that the maximum (or minimum) corresponds to the rotation velocity.
It should be noted that both methods work only when the rotation axis 
	is not exactly aligned with the line of sight,
	and the determined rotation axis and amplitude are projected quantities.
A possible systematic difference between the two methods is discussed in MP17.
One thing to note is that the difference could result from different assumptions of the two methods
	on the velocity field in a given system.
Interestingly, the results from the two methods generally agree in this study.

\begin{figure*}
\center
\includegraphics[width=0.9\textwidth]{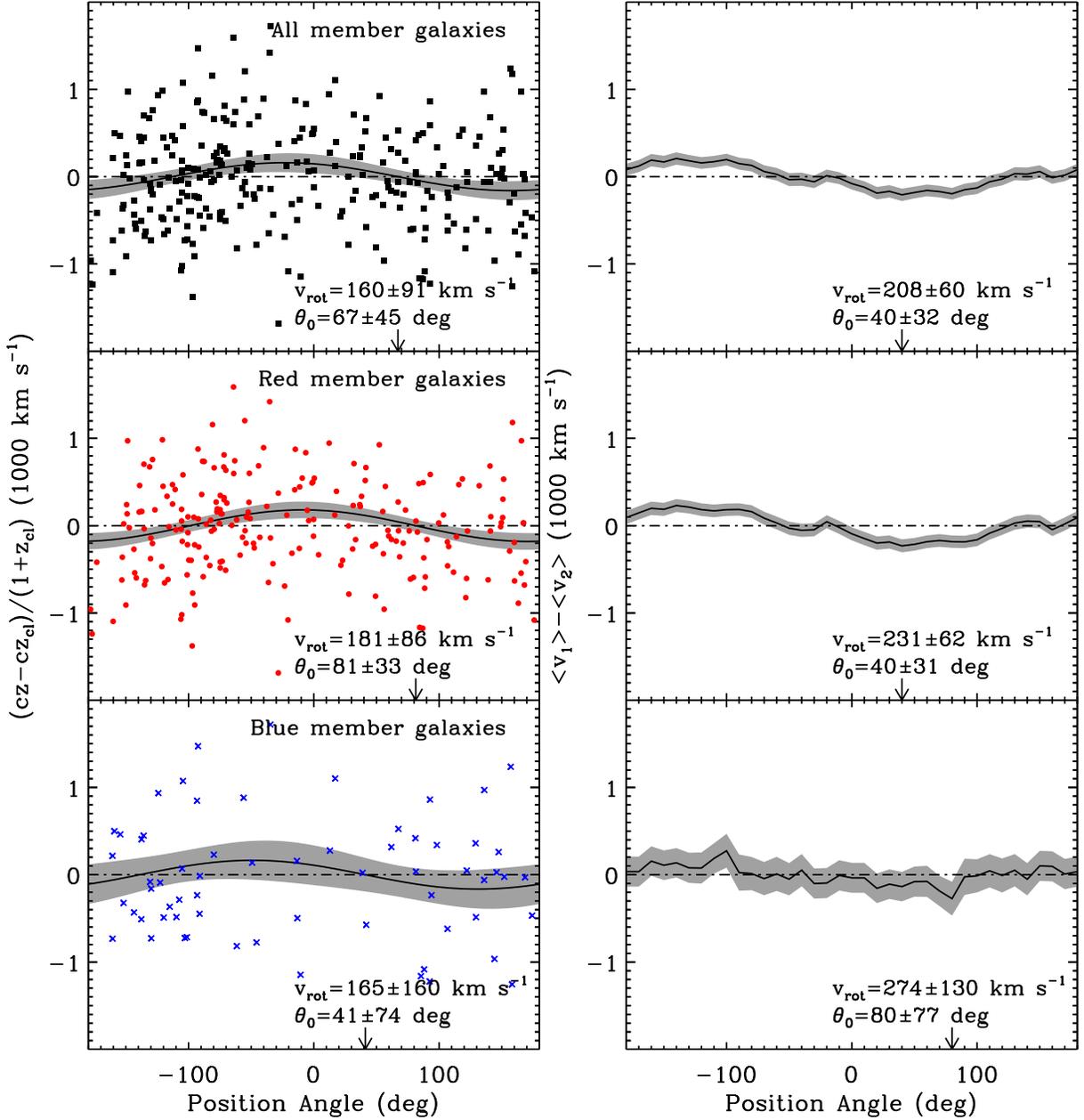}
\caption{(Left) Clustercentric velocities of the member galaxies at $R<60^\prime$
	as a function of position angle:
	black dots for all members (top), red dots for red members (middle), 
	and blue crosses for blue members (bottom).
Black solid line denotes the best-fit rotation curve for Equation (\ref{eqn-vfit}),
	and the gray band represents the 1$\sigma$ range.
(Right) Difference between two mean clustercentric velocities
	($\langle v_1\rangle$ and $\langle v_2\rangle$) of two galaxy samples 
	divided by a line of a given position angle.
Rotation velocity ($v_\textrm{rot}$) and position angle ($\theta_0$) of rotation axis 
	measured from each method for each subsample are written in each panel.
While the rotation velocity corresponds to the amplitude of the curves of both methods,
	the position angle of the rotation axis corresponds to different positions:
	the position where the sine curve intersects the X-axis from positive to negative 
	in the left panel (HL07)
	and the position where the mean velocity difference curve reaches its minimum
	in the right panel (MP17).
The position angle of rotation axis is denoted by an arrow on the X-axis in each panel.
Errors are estimated with the bootstrap resampling method.
}\label{fig-vfit}
\end{figure*}

We use both methods to determine the rotation axis and amplitude
	for all cluster members (top panels), red members (middle), and blue members (bottom) 
	at $R<60^\prime$ in Figure \ref{fig-vfit}.
The left panels show the clustercentric radial velocities of the galaxies
	as a function of position angle along with the best-fit rotation curve (black solid line) 
	for Equation (\ref{eqn-vfit}) based on the HL07 method.
The right panels show the mean velocity difference of the two galaxy samples
	as a function of position angle of the division line following the MP17 method. 
The errors are obtained through the bootstrap resampling method.
The measurements from the two methods agree with each other within 1$\sigma$ error.
Both methods suggest that cluster galaxies, mostly red galaxies, at $R<60^\prime$ 
	do have mild rotation with an amplitude of 160 or 208$\,\textrm{km s}^{-1}$
	that corresponds to $\sim28$ or $36\%$ of the velocity dispersion of 
	cluster members at $R<60^\prime$ ($\sigma_\textrm{V}=578\,\textrm{km s}^{-1}$).
This rotation is measured highly significant (i.e. 3.5$\sigma$) in the MP17 method,
	but is not so significant in the HL07 method (i.e. 1.8$\sigma$).

\begin{figure*}
\center
\includegraphics[width=0.9\textwidth]{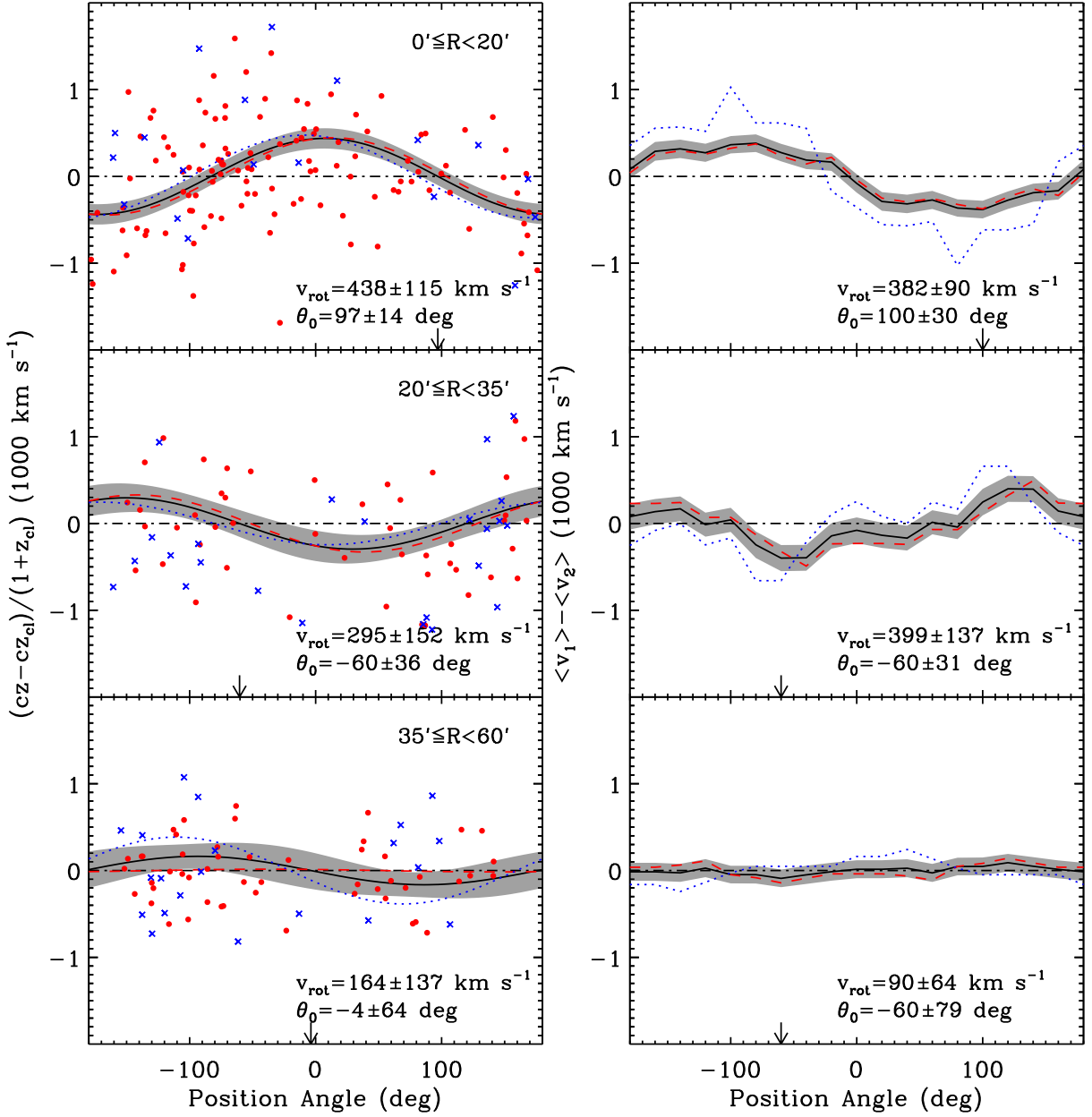}
\caption{Similar to Figure \ref{fig-vfit},
	but member galaxies in three different radial bins:
	$0^\prime \le R<20^\prime$ (top), $20^\prime \le R<35^\prime$ (middle), 
	and $35^\prime \le R<60^\prime$ (bottom).
The best-fit rotation curve (right) and $\langle v_1\rangle-\langle v_2\rangle$ curve (left) 
	are obtained for all members (black solid line with gray band),
	red members (red dashed line), and blue members (blue dotted line), respectively.
Red dots and blue crosses in the left panels represent red and blue members, respectively.
Rotation velocity ($v_\textrm{rot}$) and position angle ($\theta_0$) of rotation axis 
	measured for all members are given in numbers,
	and the position angle of rotation axis is denoted by an arrow on the X-axis in each panel.
}\label{fig-vfitr}
\end{figure*}

\begin{figure*}
\center
\includegraphics[width=0.9\textwidth]{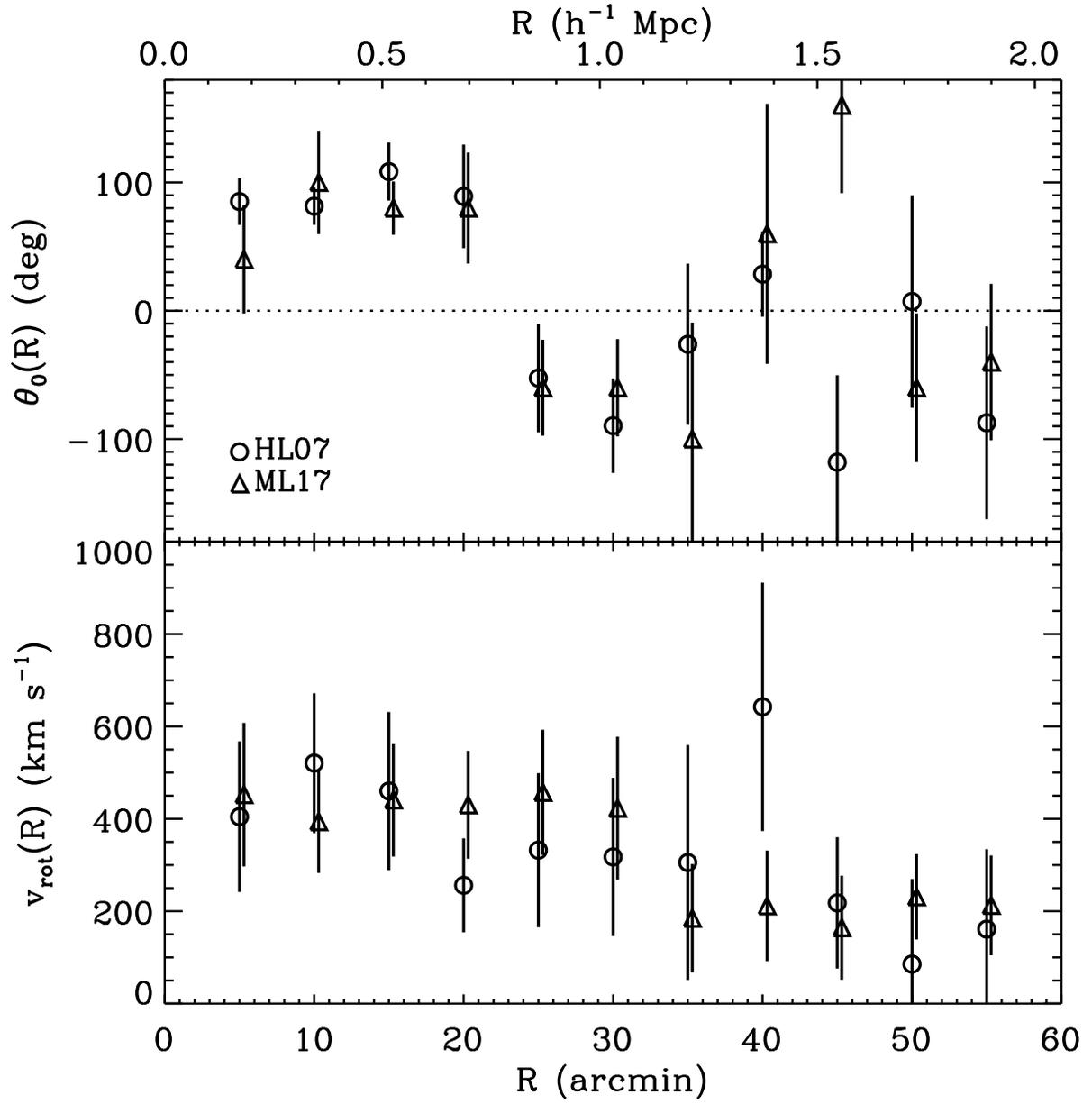}
\caption{Position angle of rotation axis ($\theta_0$, top) 
	and rotation velocity ($v_\textrm{rot}$, bottom) 
	measured from the methods by \citet[][circles]{HL2007} and \citet[][triangles]{MP2017}
	at $R<60^\prime$.
Errors are estimated with the bootstrap resampling method.
We move the triangles slightly along the X-axis to avoid the overlap with open circles.
}\label{fig-thetavrot-r}
\end{figure*}

We now examine the rotation in three different radial bins, 
	of which results are presented in Figure \ref{fig-vfitr}.
Similar to Figure \ref{fig-vfit}, 
	the left and right panels show the results for the HL07 and MP17 methods, respectively.
From top to bottom panels,
	we show the galaxies at $0^\prime<R<20^\prime$ ($0.00\,h^{-1}\textrm{Mpc}<R<0.69\,h^{-1}\textrm{Mpc}$),
	 $20^\prime<R<35^\prime$ ($0.69\,h^{-1}\textrm{Mpc}<R<1.20\,h^{-1}\textrm{Mpc}$),
	and $35^\prime<R<60^\prime$ ($1.20\,h^{-1}\textrm{Mpc}<R<2.06\,h^{-1}\textrm{Mpc}$), respectively.
We choose these radial bins to split the galaxies at $R\sim20^\prime$ and $R\sim35^\prime$
	where the rotation axis changes its direction (see Figure \ref{fig-thetavrot-r})
	and to have similar numbers of galaxies in the two outer bins (72 galaxies each).
In each panel, measurements for all, red, and blue members in a given radial bin are shown
	with black solid line with 1$\sigma$ gray band, red dashed line, and blue dotted line, respectively.
Again, the two methods give consistent results within 1$\sigma$ error.
All, red, and blue members show roughly similar motion in all radial bins.
Figure \ref{fig-vfitr} shows that there is a strong rotation signal
	at a significance of $\gtrsim4\sigma$ in the inner region (i.e. $0^\prime<R<20^\prime$)
	with an rotational amplitude of 438 or 382$\,\textrm{km s}^{-1}$,
	which accounts for 69 or 61\% of the velocity dispersion
	of cluster members in the same region ($\sigma_V=629\,\textrm{km s}^{-1}$).
The rotational signal becomes weaker with increasing clustercentric radius.
One interesting feature from this analysis is that 
	the rotation axis appears to change from the innermost region to intermediate region.
This change is clearly seen in Figure \ref{fig-thetavrot-r}, 
	which shows the position angle of rotation axis (top panel) 
	and rotation velocity (bottom panel)
	measured from the two methods (circle: HL07, triangle: MP17)
	as a function of clustercentric radius.
The results from the two methods generally agree well
	even though they differ at some large radii.\footnote{
		The rotation axis measurements from the two methods at $R\sim45^\prime$
		agree with each other within 1$\sigma$, reminding that $\theta_0=180^\circ=-180^\circ$.}
The results from the both methods show the flip of the rotation axis at $R\sim25^\prime$.
Although the rotation amplitude decreases with clustercentric radius,
	the flip of the rotation axis happens 
	where the rotation amplitude and its significance is still not low.
This can indicate a complex kinematic structure of A2107. 

To examine a possible systematics in the measurement of the rotation
	introduced by different choice of cluster center, we repeat all the measurements 
	by adopting the cD galaxy position as the cluster center
	instead of the hierarchical center position.
The derived parameters hardly change ($<1\sigma$).
We also examine the effect of the completeness of data in the rotation measurement.
To do that, we construct three different samples using the member galaxies at $R<20^\prime$
	and compare the derived parameters:
	the sample in \citet[][56 members]{OH1992} 
	and the samples before (83 members) and after (139 members) our MMT/Hectospec observations.
While the rotation axis measurements for these three samples agree with each other within $1\sigma$,
	the rotation velocity measurements are slightly different:
	$643\pm117\,\textrm{km s}^{-1}$, $432\pm119\,\textrm{km s}^{-1}$, and $438\pm115\,\textrm{km s}^{-1}$
	by the HL07 method,
	and $769\pm140\,\textrm{km s}^{-1}$, $492\pm112\,\textrm{km s}^{-1}$, and $427\pm94\,\textrm{km s}^{-1}$
	by the MP17 method.
The inclusion of the SDSS/MMT data makes a non-negligible difference in the rotation velocity measurement.
Because the completeness of the data is mainly set by the magnitude limit,
	such a difference could be caused by the luminosity (or mass) dependence of rotational motion.
To test this idea, we divide the sample at $R<20^\prime$ into two luminosity subsamples with the same size,
	and determine the rotation parameters separately.
We obtain 
	$492\pm122\,\textrm{km s}^{-1}$ (HL07) or $508\pm141\,\textrm{km s}^{-1}$ (MP17)
	for the luminous sample (70 members),
	and $162\pm223\,\textrm{km s}^{-1}$ (HL07) or $353\pm127\,\textrm{km s}^{-1}$ (MP17) 
	for the faint sample (69 members),
	indicating that the global rotation of A2107 is dominated by luminous galaxies.
This also indicates that the dependence of rotational motion on the data completeness
	does come from the luminosity dependence of rotational motion.

\section{DISCUSSION: ORIGIN OF ROTATION}\label{sec-discuss}

In this section, we test the scenario in which infall of galaxies from
	surrounding large-scale structures gives rise to the global rotation of A2107.
In Section \ref{sec-lss}, we find the large-scale structures around the cluster
	and examine their connection to the cluster.
In Section \ref{sec-ps}, we study the correlation between 
	the strength of rotational motion of cluster members
	and their infall time to the cluster potential well.

\subsection{Connection to the Large-scale Structures in the Universe}\label{sec-lss}

\begin{figure*}
\center
\includegraphics[width=0.9\textwidth]{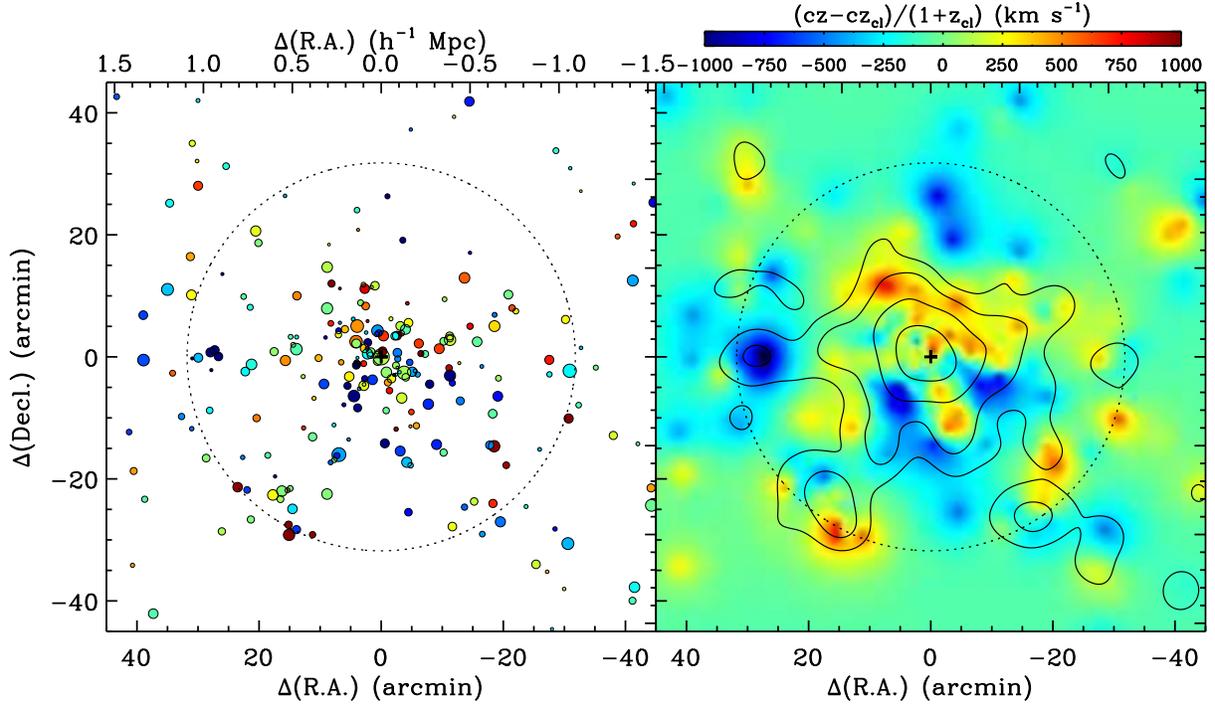}
\caption{(Left) Spatial distribution of member galaxies 
	along with clustercentric velocity information.
Brightness and clustercentric velocity of galaxies are denoted by 
	the size and color of circles, respectively.
(Right) Smoothed velocity map with number-density contours.
Density contours are equi-spaced in log by $\log 2$.
Large circle with dotted line represents $r_{200}$ of A2107
	determined from the caustic method.
}\label{fig-vmap}
\end{figure*}

\citet{OH1992} examined several possibilities that could explain such rotation-like velocity structure of A2107:
	a superposition of two clusters and a genuinely rotating cluster.
Although it is not possible to totally rule out the former case (i.e. a superposition of two clusters) 
	based only on optical spectroscopic data,
	the Gaussian velocity histogram with a single peak and the modest velocity dispersion 
	of A2107 (see the bottom right panel of Figure \ref{fig-caustic}) 
	tend to support the idea of a single cluster in rotation 
	rather than a system with two clusters .
Figure \ref{fig-vmap} shows the spatial distributions of cluster member galaxies 
	along with their velocity information.
In the left panel each circle denotes each member galaxy,
	of which size and color represent brightness (larger, then brighter) 
	and line-of-sight velocity, respectively.
The right panel shows the number-density contours on top of the smoothed velocity map.
The plot shows a clear difference in velocity structure between northern and southern parts at $R<20^\prime$,
	again indicating global rotation of the main clump.
The number density contours show that there is only one main clump at $R<20^\prime$,
	which supports the idea of a single cluster in rotation.
In addition, the second brightest member is fainter than the cD galaxy
	by more than one magnitude,
	which suggests that there is no dominant substructure.

\begin{figure*}
\center
\includegraphics[width=0.9\textwidth]{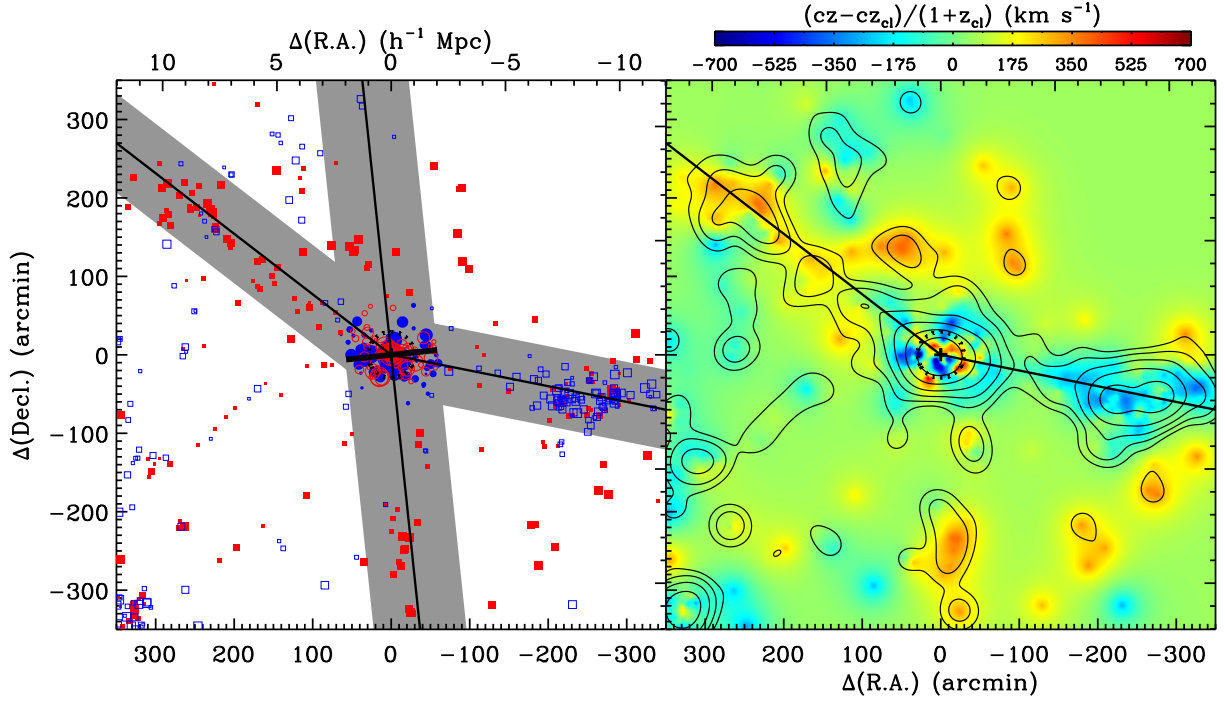}
\caption{Similar to Figure \ref{fig-vmap}, 
	but for galaxies within a redshift slice of 
	$|(cz-cz_\textrm{cl})/(1+z_\textrm{cl})|<500\,\textrm{km s}^{-1}$ 
	with A2107 in the middle.
Note its wider field of view than that in Figure \ref{fig-vmap}.
The circles with dotted line are the same as in Figure \ref{fig-vmap}, 
	representing $r_{200}$ of the cluster.
In the left panel, the member galaxies at $R\lesssim2\,r_{200}$ are denoted by circles,
	while non-members are presented with squares.
Color and size of the symbols denote the sign (red for positive; blue for negative) 
	and magnitude of clustercentric velocities, respectively.
In the right panel, number-density contours (equi-spaced in log) 
	show the connection between A2107 
	and two nearby structures in the northeast and west of the cluster.
Two thick solid lines shown in both panels connect
	the number-density peaks of the structures to the cluster center. 
Gray bands in the left panel denote the distance perpendicular to the lines, $|d|<50^\prime$.
This solid line in the north-south direction
	represents a line perpendicular to the rotation axis determined 
	for the member galaxies at $R<20^\prime$ (short very thick solid line)
	with a gray band of $4\,r_{200}$-wide.
}\label{fig-vmap-wide}
\end{figure*}

We now examine the connection of the angular momentum of A2107 
	to the surrounding large-scale structures.
We first search for large-scale structures on the sky 
	and then those along the line-of-sight direction.
Similar to Figure \ref{fig-vmap}, we show the spatial distributions of galaxies around A2107,
	but for a much larger field of view
	by including background and foreground galaxies (squares in the left panel)
	with $|(cz-cz_\textrm{cl})/(1+z_\textrm{cl})|<500\,\textrm{km s}^{-1}$ in Figure \ref{fig-vmap-wide}.
We consider galaxies that are within the caustics and at $R<2r_{200}$ as cluster members 
	(circles in the left panel).
It is noted that non-member galaxies from the SDSS DR12 have a magnitude limit
	of $m_{r\textrm{,Petro,0}}\le17.77$ 
	(spectroscopic survey completeness $\sim95\%$, \citealt{Strauss_etal2002}),
	and thus the same magnitude cut is applied to our MMT/Hectospec data for this analysis.\footnote{
		The magnitude limit used in Figure \ref{fig-vmap} is $m_{r\textrm{,Petro,0}}\le19.1$.}
In the left panel, the color and size of symbols denote the sign and magnitude of velocities, respectively.
The number-density contours in the right panel show two over-densities (structures, hereafter) 
	in the northeast and west that are connected to the cluster.
We connect each density peak of the two structures to the cluster center 
	with a thick solid line in both panels.
The regions close to the solid lines (i.e. distance perpendicular to the solid lines, $|d|<50^\prime$)
	are denoted by gray bands to indicate roughly the size of the structures.

\begin{figure*}
\center
\includegraphics[width=0.9\textwidth]{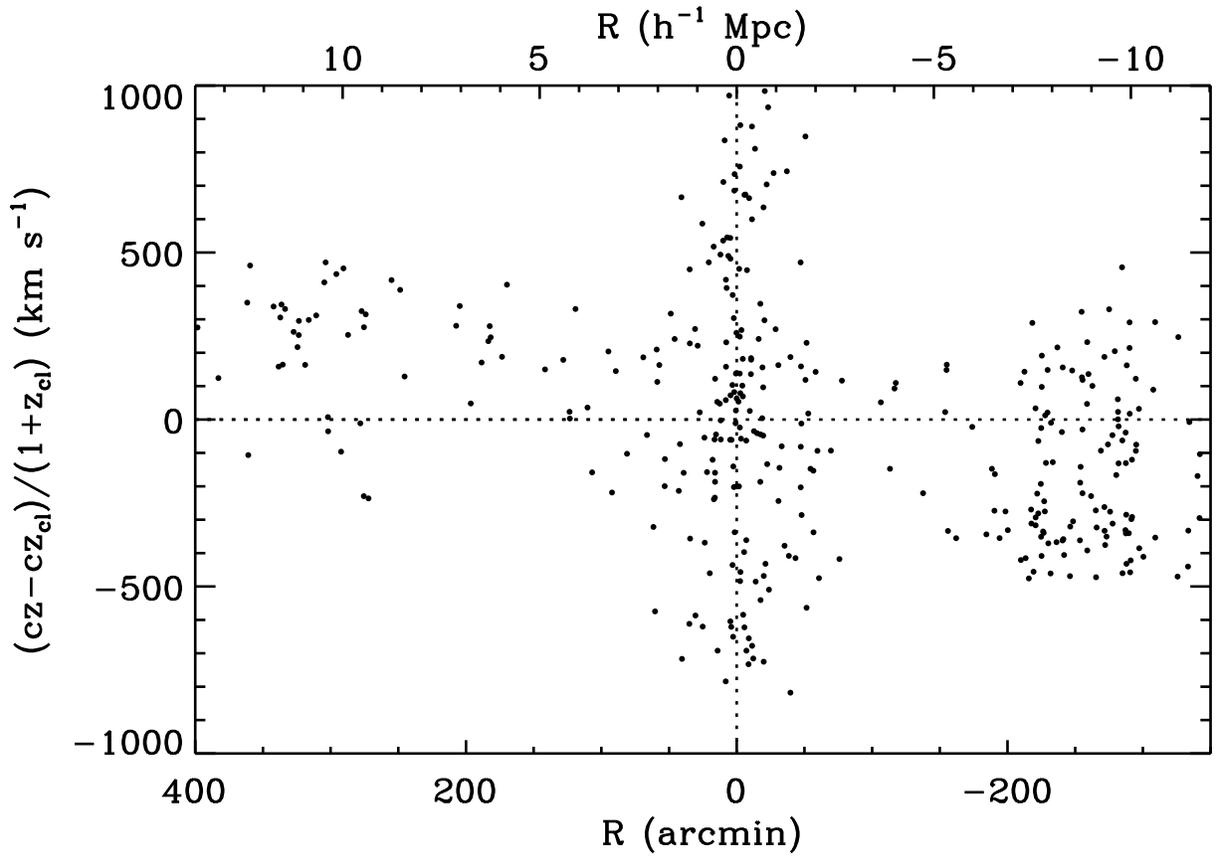}
\caption{Clustercentric velocities of galaxies
	in the northeast-west gray bands of Figure \ref{fig-vmap-wide}
	as a function of distance along the gray bands from the cluster center.
Note that groups of dots at $R\sim300^\prime$ and $R\sim-260^\prime$ 
	are the large-scale structures
	in the northeast and west of the cluster, respectively,
	which appear in the density contours in the right panel of Figure \ref{fig-vmap-wide}.
}\label{fig-Rv}
\end{figure*}

To examine how galaxies in these two structures are connected to the cluster galaxies,
	we plot the distribution of clustercentric velocities of galaxies in the gray bands in the northeast and west
	as a function of the distance along the gray bands from the cluster center in Figure \ref{fig-Rv}
	(i.e. $R$-$v$ diagram).
The distance is taken to be positive toward the east.
It should be noted that the radial velocity obtained from a redshift
	consists of two components of a receding motion (from us) driven by the cosmic expansion 
	and a peculiar motion driven by local gravitational interactions.
For galaxies at $R\lesssim30^\prime$ (inside the virial radius of A2107),
	the velocity is dominated by the peculiar motion in the cluster potential well.
For galaxies at $R\gtrsim60^\prime$ (outside the cluster potential well),
	the redshift is mainly contributed by the cosmic expansion,
	which can be considered as a distance from us.
Therefore, the over-dense region (at $R\sim300^\prime$) is probably
	a structure in the background of the cluster 
	with a receding velocity from the cluster center of $\sim300\,\textrm{km s}^{-1}$.
The other one in the west (at $R\sim-260^\prime$) is spread quite broadly along the line-of-sight direction.
Nevertheless, galaxies seem to form a structure at a distance similar to the northeast one,
	but in the foreground of the cluster.
Because the velocity changes smoothly from each of the two large-scale structures to the cluster,
	the large-scale structures appear to be connected to the cluster.
From these connections, we can infer the impact of the large-scale structures
	on the kinematics of galaxies in and around the cluster;
	galaxies infalling from the structure in the northeast (one behind the cluster) 
	would have a velocity component toward us,
	while galaxies from the structure in the west (one in front of the cluster)
	would have a velocity component away from us.
The right panel of Figure \ref{fig-vmap-wide} shows 
	such a blue blob in the east and a red one in the west 
	where the lines connecting the structures and cluster (thick solid lines) 
	cross the virial radius (dotted circle).
These blobs do not seem to directly contribute to the rotational motion detected at $R<20^\prime$, 
	but it might be responsible for the flip of the rotation axis at $R\sim20^\prime$
	(see the top panel of Figure \ref{fig-thetavrot-r}).

\begin{figure*}
\center
\includegraphics[width=0.9\textwidth]{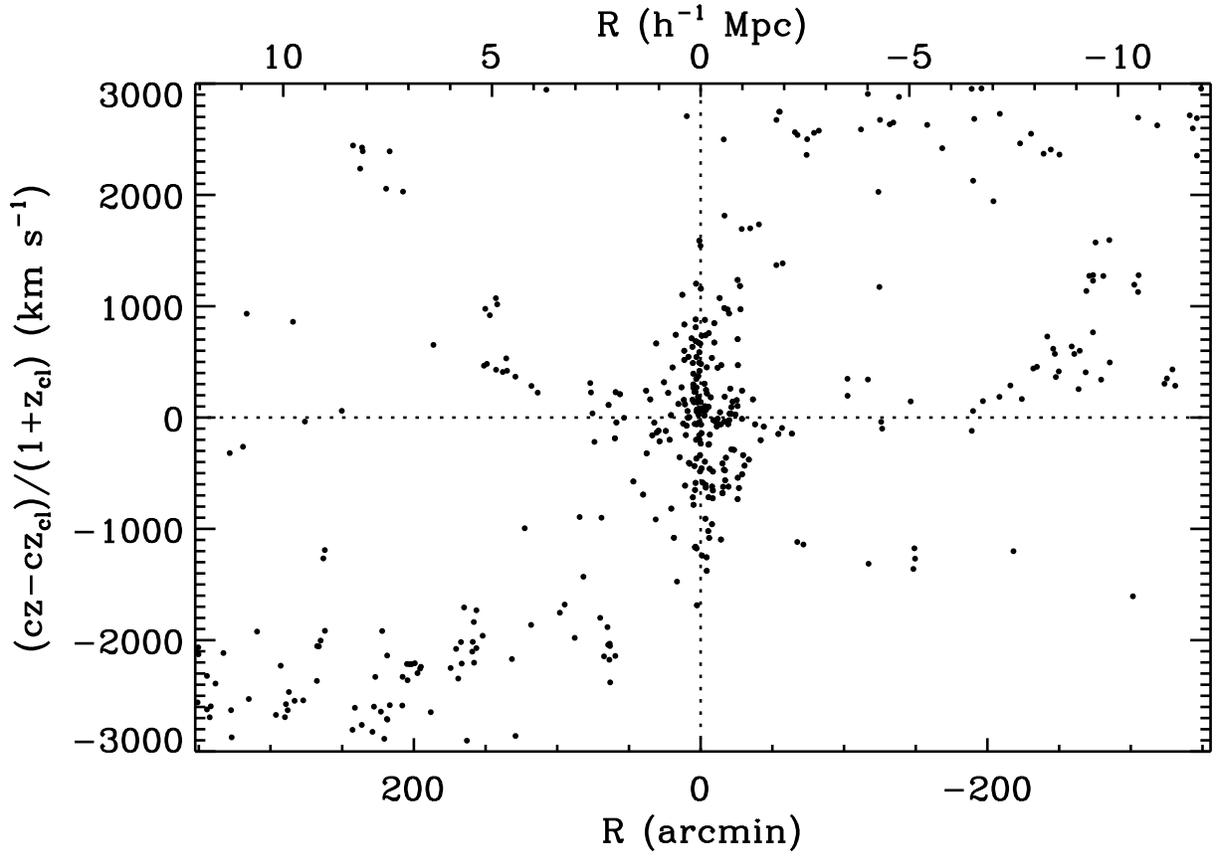}
\caption{Similar to Figure \ref{fig-Rv}, but for galaxies in the north-south gray bands
	in the left panel of Figure \ref{fig-vmap-wide}.
Note that the redshift range (i.e. Y-axis) is not restricted, unlike in Figure \ref{fig-Rv},
	to search for structures aligned along the line-of-sight direction.
}\label{fig-Rv-2}
\end{figure*}

If the global rotation of A2107 is induced by the infall of galaxies from surrounding large-scale structures,
	one might expect a large-scale structure along the direction perpendicular to the rotation axis.
Because there is no such structure in the plane of sky as seen in the right panel of Figure \ref{fig-vmap-wide}
	(note that the plot is made with the galaxies in a narrow radial velocity range),
	we search for the one that is aligned along the line-of-sight direction
	by drawing another $R$-$v$ diagram.
In the left panel of Figure \ref{fig-vmap-wide}, 
	the short thick solid line denotes the rotation axis obtained from the member galaxies at $R<20^\prime$,
	and the long thick solid line in the north-south direction is perpendicular to the rotation axis
	with a gray band of $4r_{200}$-wide.
Similar to Figure \ref{fig-Rv}, we show the distribution of the clustercentric velocities
	of galaxies in the north-south gray band 
	as a function of the distance from the cluster center in Figure \ref{fig-Rv-2}.
It should be noted that no redshift cut is imposed in Figure \ref{fig-Rv-2} unlike in Figure \ref{fig-Rv}.
Most galaxies in the north ($R>0^\prime$) have negative velocities of $\sim-2500\,\textrm{km s}^{-1}$,
	but galaxies in the south ($R<0^\prime$) form two branches, 
	one of which goes to $\sim2700\,\textrm{km s}^{-1}$, and the other goes to $\sim400\,\textrm{km s}^{-1}$ 
	(this structure appears in Figure \ref{fig-vmap-wide} as a group of red squares and a red clump in the south
	in the left and right panels, respectively).
If galaxies came into the cluster through these structures,
	galaxies from the northern structure that is in front of the cluster 
	would have a velocity component of receding from us,
	which could give rise to the same directional motion of cluster galaxies in the northern part of the cluster
	(appearing red in the right panels of Figures \ref{fig-vmap} and \ref{fig-vmap-wide}).
Similarly, galaxies from the southern structure that is behind the cluster 
	would have a velocity component of approaching toward us,
	resulting in the blueshifted motion, which is found in the southern part of the cluster.
This means that the structures in north and south 
	can naturally explain the rotational motion of the cluster members at $R<20^\prime$.
However, unlike those in the northeast and west, 
	the clustering of galaxies in these structures and their connections to A2107 seem rather weak.
Nevertheless, the anisotropic distribution of the structures along the line-of-sight direction
	(few galaxies at positive velocities in the north and at negative velocities in the south)
	and the smooth connection between neighboring and cluster galaxies in the phase space
	suggest a close link between the inflow of galaxies from the large-scale structures
	and the angular momentum of A2107.

When we use the comoving distances of galaxies derived from their redshifts
	with simple geometric relations between observed quantities
	based on the assumption that the observed motion of galaxies is pure Hubble flow,
	we can obtain the angle between the line-of-sight direction of the cluster
	and the line-of-sight direction of each structure viewed from the cluster:
	$80^\circ$ for the structure in the northeast,
	$100^\circ$ for the one in the west,
	$20^\circ$ for the one in the north,	
	$15^\circ$ for one of the two branches in the south, and
	$80^\circ$ for the other branch in the south.
This indicates that the structures in the northeast, west, and south (one of the two branches) are roughly aligned with
	the direction perpendicular to the line of sight,
	and the structures in the north and south (the other branch) are roughly aligned with the line of sight.

To identify any known galaxy groups that comprise the large-scale structures connected to A2107,
	we search for their counterparts in the SDSS galaxy group catalog
	by \citet{Tempel_etal2014}.
The details about the method, data reduction, group finding procedure
	are described in \citet{Tempel_etal2014}.
Because A2107 is a member of the Hercules supercluster 
	(\citealt{Einasto_etal2001}, Group 370 in \citet{Tempel_etal2014} catalog),
	its neighboring structures are expected 
	to be parts of the Hercules supercluster as well.
The comparison of our galaxy catalog with the group catalog of \citet{Tempel_etal2014}
	reveals that the structure in the west corresponds to 
	Groups 554 (70 members) and 1610 (62 members) 
	of \citet{Tempel_etal2014}.
There is no counterpart for the other three structures,
	when we limit groups to those with 10 members or more.
However, when we use all groups with 3 members or more,
	the remaining three structures also have their counterparts.

\subsection{Rotational Motion and the Phase-space Diagram of Cluster Galaxies}\label{sec-ps}
In the previous section, we have found large-scale structure candidates 
	that might be responsible for the global rotation of A2107.
However, it is impossible to directly capture the inflow of galaxies from the large-scale structures
	into the clusters without simulation data.
Instead, we can examine the rotational motion of cluster members 
	that have different infall histories
	to further test the idea of infall induced global rotation of clusters.
If the rotational motion of galaxies in A2107 is induced by 
	the infall of galaxies from nearby large-scale structures,
	it is expected that the galaxies that have recently come into the cluster
	(retaining more memory of large-scale structures)
	would exhibit such rotational motion more prominently than 
	those that have come earlier.
We therefore examine rotational motion 
	for subsamples of cluster members divided by their infall time.

\begin{figure*}
\center
\includegraphics[width=0.9\textwidth]{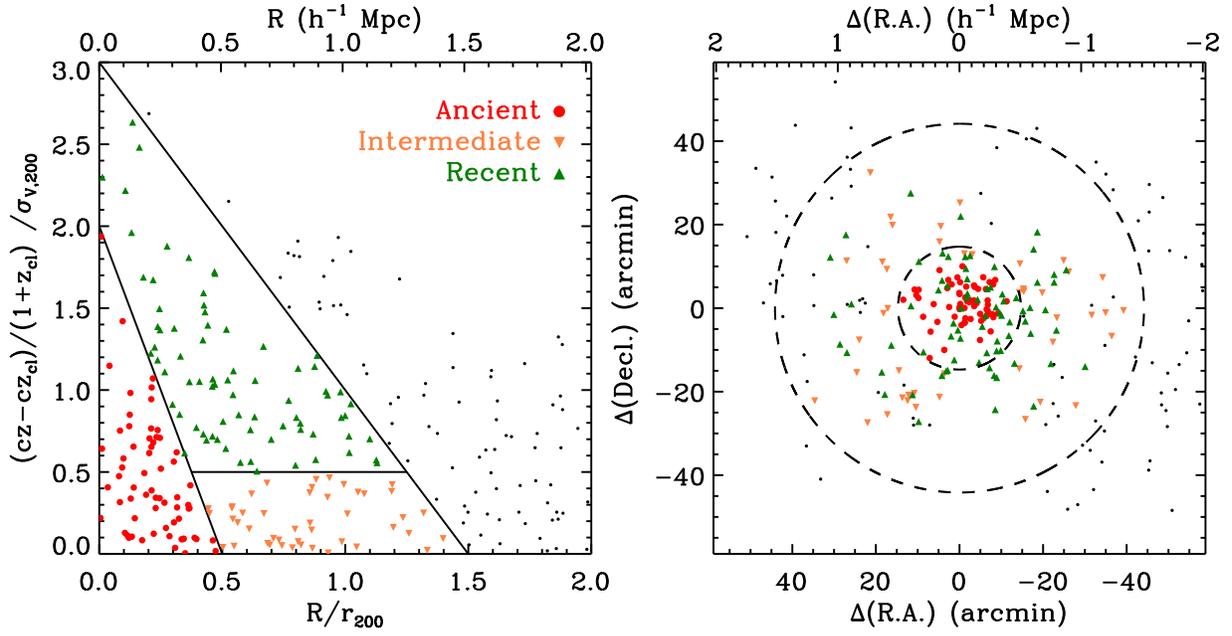}
\caption{(Left) Phase-space diagram.
Clustercentric radius and velocity are normalized by
	$r_{200}$ and the velocity dispersion of members at $R<r_{200}$ ($\sigma_{V,200}$).
Galaxies are divided into four subsamples, 
	following the classification introduced by \citet{Rhee_etal2017}:
	red dots correspond to region E in their Figure 6 (dominated by ancient infallers), 
	green triangles to regions B \& C combined (recent infallers),
	orange triangles to region D (intermediate infallers), 
	and black dots to region A (first infallers).
(Right) Spatial distribution of each subsample.
Circles with dashed lines denote $0.5\,r_{200}$ and $1.5\,r_{200}$ radii, respectively.
}\label{fig-phasespace}
\end{figure*}

The phase-space diagram (i.e. velocity-radius diagram) for galaxy clusters
	is useful for understanding their formation history 
	\citep{Gill_etal2005,Mahajan_etal2011,Oman_etal2013,Hernandez-Fernandez_etal2014,Muzzin_etal2014,
		Haines_etal2015,Jaffe_etal2015,Smith_etal2015,Jaffe_etal2016,Oman_Hudson2016}.
Recently, \citet{Rhee_etal2017} performed phase-space analyses in group and cluster environments
	to explore the evolution of galaxies since their infall to these environments 
	using the Yonsei Zoom-in Cluster Simulation data \citep{CY2017}.
They found that galaxies tend to follow a certain path in the phase space 
	as galaxies get settled in a cluster potential.
Thus the position of galaxies in the phase space can be used 
	to estimate the time since their first infall into the potential in a statistical sense.
Following their analysis, we divide the phase space of A2107 into the regions dominated by
	ancient (region E in their Figure 6), intermediate (region D), and recent infallers (regions B and C combined),
	which are represented by red dots, orange, and green triangles, respectively, 
	in the left panel of Figure \ref{fig-phasespace}.
Black dots are the galaxies that are relatively loosely associated in the cluster potential.
It should be noted that each region has a broad distribution of infall time.
However, the mean infall time changes systematically from E to B,
	so we could investigate a connection between a physical quantity and infall time
	by probing a change of the quantity over these regions.
The right panel shows the spatial distribution of these galaxies on the sky.
Neither particular elongation nor segregation is found
	except the radial segregations that are expected from the way we divide the galaxy sample.

\begin{figure*}
\center
\includegraphics[width=0.9\textwidth]{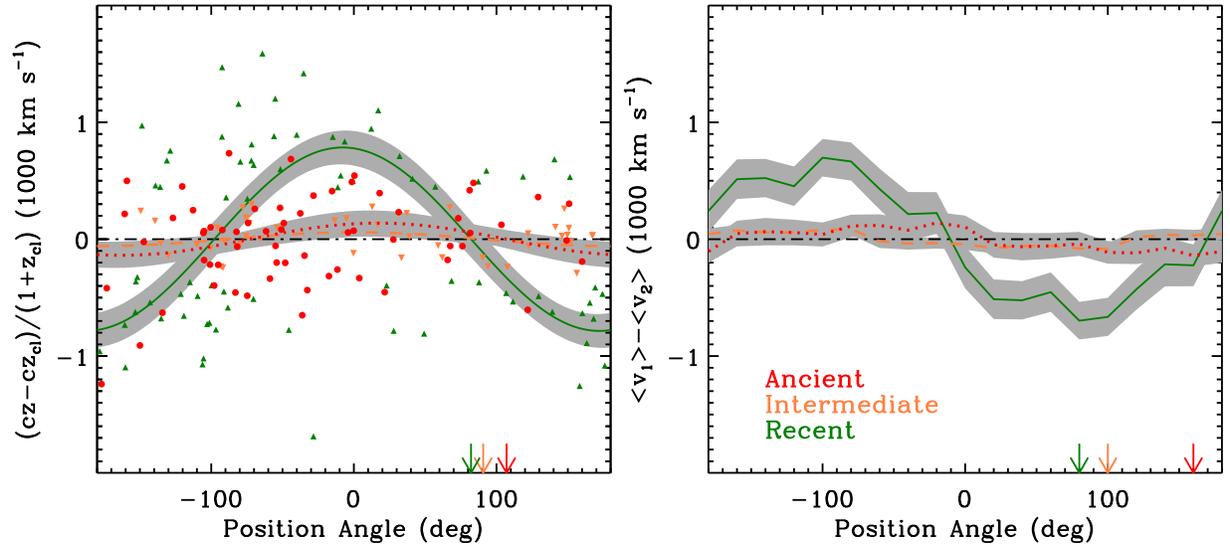}
\caption{Rotation diagrams of the methods by \citet[][left panel]{HL2007}
	and \citet[][right panel]{MP2017}
	for ancient infallers (red: dotted lines or dots) 
	and intermediate infallers (orange: dashed lines or upside-down triangles),
	and recent infallers (green: solid lines or triangles).
The position angle of rotation axis measured for each infaller sample
	is denoted by an arrow on the X-axis in its assigned color.
}\label{fig-vfit-pop}
\end{figure*}

We now determine rotation velocity and axis separately 
	for the ancient, intermediate, and recent infallers 
	using the two methods by HL07 and MP17.
The left and right panels of Figure \ref{fig-vfit-pop} show 
	the rotation diagrams of the HL07 and MP17 methods, respectively,
	similar to Figures \ref{fig-vfit} and \ref{fig-vfitr}.
Here, red, orange, and green colors represent 
	the ancient, intermediate, and recent infaller cases, respectively.
The rotation velocity measured for the ancient infallers
	are $138\pm96\,\textrm{km s}^{-1}$ by the HL07 method
	and $140\pm72\,\textrm{km s}^{-1}$ by the MP17 method.
None of the two methods find strong evidence of rotation 
	($1.4\sigma$ and $1.9\sigma$) from the ancient infallers.
The intermediate infallers have a rotation velocity of 
	$59\pm33\,\textrm{km s}^{-1}$ (HL07) or $84\pm36\,\textrm{km s}^{-1}$ (MP17).
Because of the velocity limit imposed to the intermediate infallers by definition,
	the rotation velocity of them can not be larger than that of the ancient infallers.\footnote{
		To be free from the systematic bias in the rotation measurement
		introduced by the different velocity limits 
		imposed on the ancient and intermediate infallers, 
		we construct two samples to compare: 
		ancient infallers and intermediate/recent infallers.
		The rotation amplitude for the intermediate/recent infallers
		is measured as $409\pm116\textrm{km s}^{-1}$ by the HL07 method
		and $466\pm93\textrm{km s}^{-1}$ by the MP17 method,
		which suggests that intermediate/recent infallers show 
		stronger rotational motion than the ancient infallers do.}
However, it should be noted that the significance of the measurement is higher
	for the intermediate infallers (i.e. $1.8\sigma$ and $2.3\sigma$).
The recent infallers show the strongest rotational motion
	in terms of the rotation amplitude and its significance:
	$785\pm143\,\textrm{km s}^{-1}$ with $5.5\sigma$ (HL07) 
	and $697\pm145\,\textrm{km s}^{-1}$ with $4.8\sigma$ (MP17).
The rotation axis measurements of the three groups agree with each other within $1\sigma$.
The measurement uncertainty of the rotation axis becomes smaller from the ancient, 
	through intermediate, to recent infallers:
	$107\pm48^\circ$ and $160\pm68^\circ$
	versus $91\pm44^\circ$ and $100\pm54^\circ$
	versus $82\pm12^\circ$ and $80\pm25^\circ$.
All the measurements of the rotation velocity and axis imply that
	the more recent infallers galaxies are, the stronger rotational motion they show,
	which is consistent with our expectation.

\section{CONCLUSIONS}\label{sec-summary}

We conduct a redshift survey in the field of A2107 at $R<60^\prime$ using MMT/Hectospec.
We combine 978 new redshifts measured in this study with those in the literature
	to construct an extensive catalog of 1968 galaxies in that large area.
The resulting spectroscopic completeness is high 
	(80\% at $m_{r\textrm{,Petro,0}}<19.1$) and uniform.
By applying the caustic technique to the redshift data,
	we identify 285 member galaxies, which is more than doubled
	compared to the number of members in previous studies.
Our primary results are as follow.
\begin{itemize}
\item We confirm the spatial segregation of radial velocities
	of member galaxies, suggesting global rotation of A2107.
	This is consistent with the results in previous studies.
\item We quantify the global rotation using two independent methods
	by \citet{HL2007} and \citet{MP2017}.
	Both methods suggest that there is strong rotation of 380--440$\,\textrm{km s}^{-1}$
	with 4$\sigma$ significance at small radii of $R<20^\prime$.
	The rotation amplitude slowly decreases with clustercentric radius,
	and the position angle of the rotation axis changes abruptly at $R\sim20^\prime$.
\item We find five large-scale structure candidates 
	that are connected to A2107 and consist of known galaxy groups: 
	background structures in the northeast and south, 
	and foreground structures in the west and north.
	The structures in the north and south seem responsible	
	for the rotational motion of the member galaxies at $R<20^\prime$,
	and the structures in the northeast and west seem relevant
	to the blue and red blobs in the east and west at $R\sim r_{200}$.
\item We classify the member galaxies into the ancient, intermediate, and recent infallers
	based on their positions in the phase space.
	Galaxies that more recently infall
	show more significant rotational signals.
\end{itemize}

These results can suggest that the rotational motion detected in A2107
	originates from its surrounding large-scale structures.
To further explore the co-evolution of clusters and large-scale structures,
	we plan to use cosmological simulations.
In addition, deep and wide-field X-ray data of A2107 will be helpful
	for better understanding the merging history of the cluster
	to test the scenario if off-axis merging induced the global rotation of the cluster.

\facility{MMT Hectospec}

\acknowledgments
We thank the referee for constructive comments.
We thank Woong-Tae Kim for helpful discussion on dynamics of galaxy clusters.
ME was supported by the ETAG project 
	IUT26-2, and by the European Structural Funds
	grant for the Centre of Excellence ``Dark Matter in (Astro)particle Physics and
	Cosmology" TK133.
Funding for SDSS-III has been provided by the Alfred P. Sloan Foundation, the Participating Institutions, the National Science Foundation, and the U.S. Department of Energy Office of Science. The SDSS-III web site is http://www.sdss3.org/.
SDSS-III is managed by the Astrophysical Research Consortium for the Participating Institutions of the SDSS-III Collaboration including the University of Arizona, the Brazilian Participation Group, Brookhaven National Laboratory, Carnegie Mellon University, University of Florida, the French Participation Group, the German Participation Group, Harvard University, the Instituto de Astrofisica de Canarias, the Michigan State/Notre Dame/JINA Participation Group, Johns Hopkins University, Lawrence Berkeley National Laboratory, Max Planck Institute for Astrophysics, Max Planck Institute for Extraterrestrial Physics, New Mexico State University, New York University, Ohio State University, Pennsylvania State University, University of Portsmouth, Princeton University, the Spanish Participation Group, University of Tokyo, University of Utah, Vanderbilt University, University of Virginia, University of Washington, and Yale University.

\clearpage
\begin{deluxetable}{cccrccc}
\tabletypesize{\footnotesize}
\tablewidth{0pc}
\tablecaption{MMT/Hectospec observations of the A2107 field\label{tab-obs}}
\tablehead{
\multirow{2}{*}{Field ID} & R.A. & Decl. & 
\multirow{2}{*}{Observing date} & Exposure & 
Number of & 
Number of \\ 
 & (deg, J2000) & (deg, J2000) & & (minutes) & targets & measured redshifts
}
\startdata
   a2107b15\_1 & 235.32063 & 22.071869 &   May 25, 2015 & 3$\times$15 & 256 & 246 \\
   a2107b15\_2 & 234.53011 & 21.450384 &   May 25, 2015 & 3$\times$15 & 257 & 250 \\
   a2107b15\_3 & 235.28850 & 21.498020 &  June 21, 2015 & 3$\times$15 & 255 & 231 \\
   a2107c16\_4 & 234.55217 & 22.073997 & April 16, 2016 & 3$\times$20 & 255 & 251 \\
\enddata
\end{deluxetable}

\begin{deluxetable}{rcccccrcc}
\tabletypesize{\footnotesize}
\tablewidth{0pc}
\tablecaption{Redshifts in the field of A2107 within $1^\circ$ from the cluster center\label{tab-cat}}
\tablehead{
\multicolumn{1}{c}{\multirow{2}{*}{ID}} & \multirow{2}{*}{SDSS DR12 PhotObjID} & R.A. & Decl. & $m_{r\textrm{\fns ,Petro,0}}$ & \multirow{2}{*}{probPSF\tablenotemark{a}} & \multicolumn{1}{c}{\multirow{2}{*}{$z$}} & \multirow{2}{*}{$z$ source\tablenotemark{b}} & \multirow{2}{*}{Member\tablenotemark{c}} \\ & & (deg, J2000) & (deg, J2000) & (mag) & & & & 
}
\startdata
   1 &    1237665534928355429 & 233.859078 &  21.624812 &  17.080 &  1 & $-0.00002\pm0.00001$ &  2 &  0 \\
   2 &    1237665534928356043 & 233.861980 &  21.670568 &  20.142 &  0 & $ 0.42868\pm0.00007$ &  2 &  0 \\
   3 &    1237665372257059125 & 233.869907 &  21.942797 &  18.249 &  0 & $ 0.27320\pm0.00005$ &  2 &  0 \\
   4 &    1237665372257059122 & 233.872894 &  21.984435 &  18.257 &  0 & $ 0.33219\pm0.00007$ &  2 &  0 \\
   5 &    1237665534928355532 & 233.882845 &  21.772825 &  16.265 &  0 & $ 0.04088\pm0.00001$ &  2 &  1 \\
   6 &    1237665534928355966 & 233.884793 &  21.802447 &  20.717 &  0 & $ 0.62591\pm0.00023$ &  2 &  0 \\
   7 &    1237665372257059083 & 233.901961 &  21.822620 &  16.694 &  0 & $ 0.14371\pm0.00003$ &  2 &  0 \\
   8 &    1237665372257059957 & 233.902679 &  21.926681 &  20.975 &  0 & $ 0.71415\pm0.00020$ &  2 &  0 \\
   9 &    1237665371720253605 & 233.904397 &  21.503921 &  15.010 &  0 & $ 0.04191\pm0.00002$ &  2 &  1 \\
  10 &    1237665371720253508 & 233.912087 &  21.547363 &  18.890 &  1 & $-0.00083\pm0.00003$ &  2 &  0 \\
\enddata

\tablenotetext{a}{(0) Extended source, (1) Point source.}
\tablenotetext{b}{(1) This study, (2) SDSS DR12, 
	(3) \citet{Cava_etal2009}, (4) \citet{OH1992}, (5) \citet{Smith_etal2004}, (6) \citet{White_etal2000}, 
	(7) \href{http://tdc-www.harvard.edu/cgi-bin/arc/fsearch}{FAST Spectrograph Archive}}
\tablenotetext{c}{(0) A2107 non-member, (1) A2107 member.}
\tablecomments{This table is available in its entirety 
	in a machine-readable form in the online journal.
	A portion is shown here for guidance regarding its form and content.}
\end{deluxetable}

\clearpage
\clearpage
\clearpage
\clearpage
\clearpage
\clearpage
\clearpage 
\clearpage 
\clearpage 
\clearpage 
\clearpage 
\clearpage 
\clearpage 
\clearpage 
\clearpage 
\clearpage 
\clearpage 

\clearpage
\bibliography{ms}{}
\bibliographystyle{apj}

\end{document}